\documentclass[
twocolumn,
aps,prd,
nofootinbib,
superscriptaddress,
showpacs,ligh
tightenlines, 
]{revtex4}

\usepackage{amsmath}
\usepackage{amssymb}
\usepackage{bm}
\usepackage{color,graphicx}
\usepackage{slashed}
\usepackage{graphicx}

\usepackage{mciteplus}
\usepackage[dvipsnames]{xcolor}

\newcounter{comment}

\begin{document}
\title  {Bernstein Polynomials based Probabilistic Interpretation of Quark Hadron Duality}

\author{Evan Askanazi}
\email{ema9u@virginia.edu}
\affiliation{University of Virginia - Physics Department,
382 McCormick Rd., Charlottesville, Virginia 22904 - USA} 
\altaffiliation{Present address: 
Hebrew University of Jerusalem,
Institute of Chemistry,
Edmond J Safra Campus,
Jerusalem 9190401}

\author{Simonetta Liuti }
\email{sl4y@virginia.edu}
\affiliation{University of Virginia - Physics Department,
382 McCormick Rd., Charlottesville, Virginia 22904 - USA \\ and Laboratori Nazionali di Frascati, INFN, Frascati, Italy.}

 \begin{abstract}
It is now widely recognized that large Bjorken $x$ data play an important role in global analyses of Parton Distribution Functions (PDFs) even at collider energies, through perturbative QCD evolution. 
For values of the scale of the reaction, $Q^2$, in the multi-GeV region the structure functions at large $x$ present resonance structure. Notwithstanding, these data can be incorporated in the analyses by using quark-hadron duality or approximate scaling of the structure function data averaged over their resonance structure. Several averaging methods have been proposed using either the PDFs Mellin moments, or their truncated moments. We propose an alternative method using Bernstein polynomials integrals, or Bernstein moments. Bernstein moments render a smooth form of the structure function in the resonance region. Furthermore, being based on a different averaging criterion than the methods adopted so far, they provide a new framework for understanding the possible mechanisms giving origin to the phenomenon of quark-hadron duality.     
\end{abstract}

\maketitle

\baselineskip 3.0ex
\section{Introduction}
A vast amount of data accumulated for more than three decades at high energy colliders  has allowed us to pin down with high precision the quark and gluon structure of the proton in terms of parton distribution functions
(PDFs) \cite{Cooper-Sarkar:2015boa,Forte:2013wc,pdg}. PDFs are the parton longitudinal momentum density distributions given as a function of Bjorken $x$, the longitudinal parton to proton momentum fraction, and of the scale of the high energy interaction given by $Q^2$, the square of the four-momentum transfer in Deep Inelastic Scattering (DIS).
High energy lepton-proton and hadron-hadron scattering processes provide experimental access to PDFs thanks to the validity of QCD factorization theorems  which allow us to write the measured cross section as the convolution of 
a theoretically known part describing the hard-probe target-parton scattering with linear combinations of PDFs \cite{Collins:1989gx}. PDFs are non-perturbative objects in QCD. 

The simultaneous determination of the valence, sea quark and gluon distributions from experiment is achieved through global fits including data from lepton-hadron and hadron-hadron scattering processes. Global fits have been characterized  from the inception by many nontrivial issues mostly due to the fact that 
gluons do not couple directly to the lepton probe, 
and to the consequent intricacies of Perturbative QCD (PQCD) calculations, which regulate the scale dependence of the hard scattering process 
involving  the strong coupling, $\alpha_S$, as an additional parameter to be conjointly extracted and evaluated \cite{Bethke:2015etp}.
Presently, PDFs can be determined using PQCD evolution equations at Next-to-Next-to-Next-to Leading Order (NNLO) including 
quantitative evaluations of the uncertainty of the extracted PDFs within each given framework  \cite{pdg}.

Most studies have been focused on the small $x$ region ($x \ll 0.1$) which is where high energy colliders measurements are centered. An accurate extraction of PDFs, however, requires using data in a wide range of $x$ and $Q^2$ \cite{Accardi:2016qay,Accardi:2009br}.  The large momentum fraction region, in particular, has been shown to impact the extraction of PDFs from even the highest energy colliders data through PQCD evolution \cite{Sirunyan:2017azo}.  

At large $x$ QCD factorization theorems have been  applied under the assumption that 
small-coupling techniques can be used \cite{Moffat:2017sha}. The largest  $x$ values, however, can be reached experimentally in electron proton scattering with $Q^2$ in the multi-GeV region and low final state invariant mass, $W^2$. In this kinematic
range it becomes unavoidable to consider the interplay between perturbative and non-perturbative effects which could possibly even lead to a breaking of factorization theorems.  
It is therefore interesting to carry out the extraction of PDFs at these very large $x$ values starting from the
assumption that the phenomenon of quark-hadron duality is at work when $W^2 \lesssim$ 4 GeV$^2$, and to subsequently monitor the appearance of factorization breaking effects. 

Quark-hadron duality can be viewed as a manifestation of alternating
PQCD and non-perturbative QCD (NPQCD) behavior in the proton structure. 
The appearance of quark-hadron duality in the data is characterized by the observation of an approximate $Q^2$
scaling in the region where the proton's resonance structure is clearly detectable. By ``approximate scaling" one refers to 
the similarity between the behavior at large $x$ of the  structure function's {\it average} over the resonance region at low $Q^2$, {\it i.e.} where hadron degrees of freedom (d.o.f.) dominate, 
and the large $x$, large $Q^2$ behavior which is determined by quark and gluon d.o.f..
Understanding the dynamical origin of this behavior remains one of the unsolved questions in QCD with a direct impact onto the phenomenology of the bigger and multifaceted question of monitoring the transition from its perturbative to 
non-perturbative regimes \cite{Dokshitzer:2000yu,Dokshitzer:2005bf}, including the behavior of $\alpha_S$ in the non-perturbative limit \cite{Prosperi:2006hx,Courtoy:2013qca,Deur:2016tte}.

A fundamental aspect of analyzing the large $x$ behavior of structure function at large $x$ concerns the way the average over the resonance region is computed. 
In this paper we propose a strategy, using the Bernstein polynomials technique, to average over the resonance region and we study its physical implications. 

Bernstein polynomials in $x$ are defined in terms of the power basis $\{1, x, x^2, ...x^n\}$, over the interval  $[0,1]$, so that, for a given $n$, one has $n+1$ polynomials for $k=0,...n$, characterized by one single local maximum occurring at equal spacing between $0$ and $1$ (see {\it e.g.} Ref.\cite{Bernstein} and Section \ref{sec2}).   
Because of the latter property, the Bernstein moments of order $k$ of a given function of $x$ defined in the interval $x \in [0,1]$, represent weighted averages of the function which emphasize specific ranges around the local maxima where the Bernstein $k$-polynomial is centered. As we explain later on, this will allow us to evaluate the average value of the structure function at a given $x$ with a  calculable dispersion. 

When used to approximate functions point-wise, Bernstein polynomials are characterized by a slow convergence. Notwithstanding, their most  successful application has been in computer science,
or in problems such as potential surface simulation in atomic physics \cite{HoRabitz}, {\it i.e.}  in situations for which, owing to the inherent pixelization, the knowledge of a continuous curve reconstruction is not needed, and a finite number of Bernstein polynomials can be efficiently implemented to design the parametric curves and surfaces. 

Bernstein polynomials have been used in PQCD based analyses 
of DIS data as a way to infer the $x$ dependence of the structure function knowing a finite number of its Mellin moments  \cite{Yndurain:1977wz,Pennington:1982kr,Santiago:1999pr,Santiago:2001mh}. 
More recent analyses have been extended to the DIS structure function in the electroweak sector \cite{Kataev:2001kk,Brooks:2006wh,Maxwell:2002mt}, and to Generalized Parton Distributions (GPDs)  \cite{Ahmad:2007vw}. 

In this paper, by using Bernstein moments to analyze the DIS structure functions at kinematics where the resonance spectrum is prominent, we introduce a novel perspective on the averaging procedure that might shed light on the mechanisms generating the phenomenon of quark-hadron duality. As a mathematical rendering of the phenomenon of duality, Bernstein polynomials select ranges in $x$ that are determined completely independently from those characteristic of resonance structure.
\footnote{The ranges encompassed by each Bernstein moment do not coincide at any given $Q^2$ with any of the most prominent resonance ranges.}
The resulting, reconstructed proton structure function traces a smooth function in $x$ that approximately scales with $Q^2$.
By increasing gradually the number of moments, and consequently restricting their ranges in $x$ one can define a critical number and interval size after which the smoothness of the curve is disrupted and the characteristic resonance structure starts reappearing. 
By sampling the structure function with Bernstein polynomials, one, therefore, obtains quantitative clues on the degree of ``locality" of parton-hadron duality. 
If we broadly attribute the resonance peak formation to an effect of confinement, 
we can study its emergence by making a one to one correspondence between the number of Bernstein moments used to reproduce the function's behavior in the resonance region and the number of partonic configurations generated from a probabilistic point of view. 
The occurrence of parton-hadron duality even at the local level, signals that the process is PQCD driven.  We can interpret the underlying mechanism proceeding from low to large $W^2$ as being initiated from a quark with the same $x$ value in both cases. At large $W^2$, a large number of different final state hadronic configurations is generated, each configuration having a lower probability to occur whereas, at low $W^2$, final configurations with similar hadronic content can be formed with a high probability for their occurrence.  
Bernstein polynomials allow us to describe ideally this situation.
Our approach presents similarities to parton shower Monte Carlos (MC’s) approaches \cite{Lai:2009ne,Frixione:2002ik}. The probabilistic picture behind our averaging procedure can be extended also to the recent analysis of Jefferson Lab Hall C data, where the averaging was obtained by randomly varying resonances spectra for different $Q^2$ values \cite{Christy}.   

On the practical side, our study complements previous analyses of quark-hadron duality based on Mellin moments \cite{Niculescu:2015wka,Niculescu:2000tk}, and on truncated moments of structure functions -- integrals of structure functions over restricted regions of $x$, describing local (resonance by resonance) quark-hadron duality in QCD \cite{Psaker:2008ju}. A previous analysis conducted in Ref.\cite{Bianchi:2003hi} also provided a partial reconstruction of the $x$ dependence of $F_2$. However, the average value of $F_2$ was taken over the entire resonance region for each kinematic binning in $Q^2$, yielding only one point in $x$ as the average in each interval $[x_{min}(Q^2), x_{max}(Q^2)]$ 
The analysis presented here allows us to define several points in $x$ for each $Q^2$ bin. The  smooth curve  obtained from the reconstruction procedure for $F_2$, extend to the highest experimental values of $x$. Using our averaged curves, and evaluating the impact of Target Mass Corrections (TMCs), Large $x$ Resummation (LxR) effects, and dynamical Higher Twists (HTs) would allow one to disentangle appropriately the contributions at the largest values of $x$ to be used in global fits \cite{Malace:2009kw}.

This paper is organized as follows: in Section \ref{sec2} we describe the Bernstein moments analysis and averaging procedure; in Section \ref{sec3} we present our results based on the analysis; in Section \ref{sec4} we draw conclusions and delineate our perspectives for future work. 

\section{Bernstein Moments Analysis of the electron nucleon scattering resonance region}
\label{sec2}
Reconstructing a function knowing its moments is a challenging problem from the mathematical point of view: 
imposing the necessary and sufficient conditions on the moments in order to define a unique solution to the problem renders a point-by-point reconstruction unattainable
\cite{Gaemers:1980hc}.
In physical problems, however, we deal with the {\em reduced moment problem}, where the function is reconstructed knowing only a finite number of its moments. The reconstruction can be made at the price of introducing a calculable uncertainty on both the function and the range of the variable it depends on and with respect to which the moments are calculated. 
The physics approach is, of course, consistent with the way experimental measurements are performed by presenting the observable with their uncertainties in bins of the kinematic variables they depend on.  
 
The approach that was originally introduced in Ref.\cite{Yndurain:1977wz} (see \cite{Pennington:1982kr} for a review)
to obtain the DIS structure function, $F_2$, knowing a finite number of its Bernstein moments is actually a reduced moment problem yielding $F_2$ values with a theoretical error centered in the calculated $x$ bins. Bernstein polynomials are ideal 
for reproducing the deep inelastic structure functions in that they are zero at the
endpoints, they are normalized to one, and they are peaked in different regions
of the interval $x \in [0,1]$. Because of the latter property the Bernstein
polynomials allow one to emphasize the behavior of the structure function in
specific regions of $x$, while suppressing others.  
It was found, in particular, that $n \geq 8$ moments were necessary to give a  quantitative
description of the behavior of $F_2(x,Q^2)$ in the large $x$ region consistently with the  data precision available at the time \cite{Yndurain:1977wz}. 

Our aim here is to use a Bernstein moments based reconstruction of $F_2$ in the resonance region as an averaging procedure. 
The smooth curves we obtain allow us to extend PDF fits 
of DIS data to the very large $x$ region. We provide one of such fits based on an artificial neural networks algorithm known as Self-Organizing Maps (SOM) \cite{Askanazi:2014gxa,Honkanen:2008mb}. 

\subsection{Experimental Data and the Onset of Parton-Hadron Duality}
In a QCD-based definition of quark-hadron duality one can ascribe the approximate scaling with $Q^2$ in the resonance region to cancellations among contributions of power corrections of order ${\cal O}(1/Q^2)$ and higher, of both kinematic origin and from the twist expansion, that would otherwise be expected to dominate the cross section  (see review in \cite{Melnitchouk:2005zr}).


The inclusive DIS cross section of unpolarized electrons off
an unpolarized proton is written in terms of the two structure
functions $F_2$ and $F_1$, 
\begin{widetext}
\begin{eqnarray}
\label{xsect}
 \frac{d^2\sigma}{dx dy} \equiv  
\left(\frac{4 \pi E}{x} \sin^2 \theta/2 \right)  \frac{d^2\sigma}{d\Omega d E'} =  
\frac{4\pi\alpha^2}{Q^2 xy}
\left[
    \left(1-y-\frac{(Mxy)^2}{Q^2}\right)F_2 +
    y^2 x F_1
\right] 
\end{eqnarray}
\end{widetext}
where we gave both expressions in terms of the invariants $x=Q^2/2 M \nu$, and $y=\nu/E$, and in terms of the laboratory variables $\Omega$ and $E'$, with $\nu=E-E'$, the energy transfer, $E$ and $E'$ being the initial and final electron energies, respectively, and $M$ the proton mass.
The structure functions are related by the equation,  
\begin{equation}
\label{R}
F_1 = F_2(1+\gamma^2)/(2x(1+R)),
\end{equation}
$\gamma^2=4M^2x^2/Q^2$; $R$ is ratio of the longitudinal to transverse
virtual photo-absorption cross sections.  
In QCD, $F_2$ is written as,
\begin{eqnarray}
\label{t-exp} 
F_{2}(x,Q^2) = F_{2}^{LT}(x,Q^2) +
\frac{H(x,Q^2)}{Q^2} + {\cal O}\left(1/Q^4 \right),
\end{eqnarray}
%
where $F_{2}^{LT}(x,Q^2)$ is the leading twist (LT) term, and the terms of ${\cal O} (Q^{2})$, and  higher are the genuine  
Higher Twist (HT) corrections  
that involve interactions between the struck parton and the spectators, or
multi-parton correlation functions.
The LT part depends on $Q^2$ owing to the effect of PQCD evolution,  
the finite mass of the initial nucleon (TMC), and large $x$ resummation (LxR). 

The availability of various high precision measurements \cite{Niculescu:2000tk,Malace:2009kw,Malace:2009dg,Malace:2011ad,Baillie:2011za,Tkachenko:2014byy,Tvaskis:2016uxm} has enabled detailed studies of various sources of scaling violations affecting the structure functions in addition to standard PQCD evolution, namely Target Mass Corrections (TMCs), Large x Resummation effects (LxR) and dynamical Higher Twists (HTs).
The first studies performed in Refs.\cite{Liuti:2001qk} showed the presence of non trivial QCD effects besides PQCD evolution. This was confirmed in the moments analysis of \cite{Liuti:2001qk,Ricco:1998yr,Armstrong:2001xj,Osipenko:2003bu} and again, with an increased precision in  \cite{Niculescu:2005rh,Monaghan:2012et,Psaker:2008ju,Accardi:2009br}. In Ref.\cite{Bianchi:2003hi} it was shown that LxR effects impact the onset of duality in the region $x \gtrsim 0.7$. Similar effects were found in the electroweak sector in Ref.\cite{Corcella:2005us} and in polarized semi-inclusive DIS in \cite{Anderle:2013lka}. More recent results including a thorough analysis of all PQCD generated effects have been shown in Refs.\cite{Accardi:2016qay}. The perhaps, to date, unanimous conclusion of both experimental and theoretical studies is that once PQCD evolution including LxR and TMCs are properly taken into account,  the space left for dynamical higher twist contributions can be determined precisely and it is small, at the few percent level \cite{Psaker:2008ju}. 
A smooth curve representing $F_2$ can therefore, in principle, be extracted from the resonance region that can be directly used in global PDF fits. 

Although a proper treatment of TMCs requires in principle a precise determination of the support in $x$ of the structure function \cite{Accardi:2008ne}, here we use the standard approach

\begin{eqnarray}
\label{TMC}
F_{2}^{TMC}(x,Q^2) & = &
    \frac{x^2}{\xi^2\gamma^3}F_2^{\mathrm{\infty}}(\xi,Q^2) + \\ & &
    6\frac{x^3M^2}{Q^2\gamma^4}\int_\xi^1\frac{d \xi'}{{\xi'}^2} \nonumber
F_2^{\mathrm{\infty}}(\xi',Q^2),
\end{eqnarray}
with
\begin{eqnarray}
\xi = \frac{2x}{1 + \gamma}, \quad \;\;\; \gamma = \sqrt{1 + \frac{4 x^{2} M^{2}}{Q^{2}}} .
\end{eqnarray}
$F_2^{\infty}$ is obtained from PDFs that do not contain TMCs.

LxR effects arise formally from terms containing powers of 
$\ln (1-z)$, $z$ being the longitudinal 
momentum fraction integration variable in the evolution equations, that are present in 
the Wilson coefficient functions ($z > x$). 
The latter connect the parton distributions to the observable or, in our case, 
to the structure function $F_2$, through an integral relation.
The logarithmic terms in the Wilson coefficient functions become very large at large $x$, and they need to be 
resummed to all orders in $\alpha_S$. 
As a consequence of taking into account large $x$ resummation effects, the argument of the strong coupling constant also becomes $z$-dependent, 
$\alpha_S(Q^2) \rightarrow \alpha_S(Q^2 (1-z)/z)$ \cite{Bianchi:2003hi,Courtoy:2013qca,Roberts:1999gb}. 

Perturbative QCD analyses use the 
Mellin moments of the structure function,
which allow for a more straightforward comparison with $Q^2$ dependent theoretical predictions, 
\begin{equation}
M_n(Q^2) = {\int_0^1} dx x^{n-2} F_2(x,Q^2),
\label{CN}
\end{equation}


The onset of parton-hadron duality was also studied 
by considering yet another set of integrals of the structure function \cite{Liuti:2001qk,Bianchi:2003hi},
\begin{equation}
\label{Iexp}
I^{\mathrm{res}}(Q^2) = \int^{x_{\mathrm{max}}}_{x_{\mathrm{min}}} 
F_2^{\mathrm{res}}(x,Q^2) \; dx
\end{equation}
where $F_2^{\mathrm{res}}$ is evaluated using the experimental data  
in the resonance region, and
\begin{equation}
I^{DIS}(Q^2) = \int^{x_{\mathrm{max}}}_{x_{\mathrm{min}}} F_2^{DIS}(x,Q^2) \; dx,
\label{ILT}
\end{equation}
In Eq.(\ref{Iexp}) one has for each $Q^2$ value,
\begin{subequations}
\label{eq:xminmax}
\begin{eqnarray}
x_{\mathrm{min}} & = & \frac{Q^2}{Q^2+W_{\mathrm{max}}^2-M^2}, \\
x_{\mathrm{max}} & = & \frac{Q^2}{Q^2+W_{\mathrm{min}}^2-M^2} .
\end{eqnarray}
\end{subequations}
$W_{\mathrm{min}}$  and $W_{\mathrm{max}}$ delimit the resonance region. 
Eq.(\ref{ILT}) is calculated in the same range of 
$x$ and for the same value of $Q^2$, using parametrizations 
of $F_2$ that reproduce the DIS behavior
of the data at large $Q^2$.

Duality is attained when the ratio,
\begin{equation}
R_{\mathrm{unpol}} = \frac{I^{\mathrm{res}}}{I^{DIS}},
\label{RdualLT}
\end{equation}
attains unity. 
The integral in Eq.(\ref{Iexp}) can be plotted as a function of the average value of $x$ in each interval 
$[x_{\mathrm{min}}(Q^2), x_{\mathrm{max}}(Q^2)]$. This 
was evaluated in Ref.\cite{Bianchi:2003hi} to be, $\langle x \rangle = x(W^2 \equiv 2.5 \, \rm{GeV}^2)$. 
Notice that this analysis presents similarities to the ``truncated moments" analyses of Ref.\cite{Psaker:2008ju}.  

As we explain below, Bernstein moments generalize the averages defined in Eqs.(\ref{Iexp},\ref{ILT}) 
for each $Q^2$ value, allowing for a point-wise in $x$ comparison of $F_2$ in the DIS and resonance dominated 
regions. 

A further advantage of the Bernstein polynomials based analysis is that it is much less sensitive to the elastic contribution, an issue that has been otherwise raising ambiguities in integral-based analyses.

\subsection{Bernstein Moments Averaging}
\begin{figure}[htp]
\centering{
\includegraphics[scale=0.3]{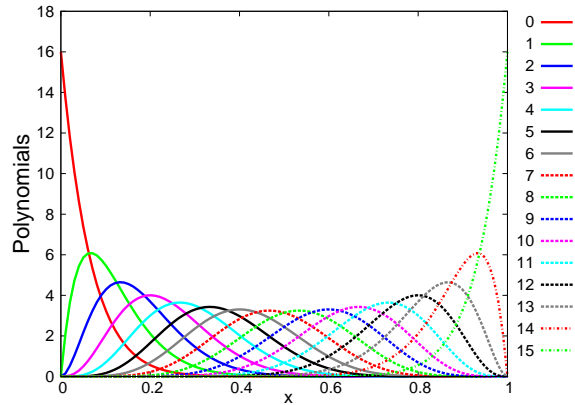}}
\caption[Bernstein Polynomials]{The Bernstein polynomials for $k=0$ to $k=15$.  The labels correspond directly to polynomial numbers.  The line with label $0$ is the polynomial for $k=0$, the line labeled $1$ is the polynomial for $k=1$ and the line labeled $2$ is the polynomial for $k=2$ going up to $k=15$}  
\label{fig:Bernpoly}
\end{figure}

The Bernstein polynomials allow us to construct a function,
$F_2^{n, k}(x)$, of which we calculate its first $n$ moments, that will converge to $F_2(x)$ for $n \rightarrow \infty$.
The polynomials functional form is given by (Figure \ref{fig:Bernpoly}),
\begin{subequations}
\begin{eqnarray}
\label{eq:171}
&& B_{n,k}(x)  = C_{n,k} \, x^{k} (1-x)^{n - k}  \quad k=0,...,n \\
&& C_{n,k}  =  \frac{\Gamma(n + 2)}{\Gamma(k + 1)\Gamma( n - k + 1)},  
\end{eqnarray}
\end{subequations}
so that the normalization condition is,
\begin{equation}
\int_0^1 \mathrm{d}x \hspace{1mm} B_{n,k}(x)  = 1. 
\label{norm}
\end{equation}
The moments of $F_{2}^{exp}$ are evaluated as, 
\begin{equation}
\label{eq:175}
F^{n,k}_2(\left\langle x  \right\rangle_{n,k}, Q^{2}) =  \int_0^1 \mathrm{d}x \hspace{1mm} B_{n,k}(x) \hspace{1mm} F_{2}^{exp}(x,Q^{2}), 
\end{equation}
where $F_2^{exp}(x,Q^2)$ is obtained directly from experiment; 
$\left\langle x  \right\rangle_{n,k}$, the average value in $x$ for each moment is, 
\begin{equation}\label{eq:173}
\left\langle x  \right\rangle_{n,k} = \int_0^1 \mathrm{d}x \hspace{1mm} x \hspace{1mm} B_{n,k}(x) = \frac{k + 1}{n + 2},
\end{equation}
Because of their shape which selects specific and sequential ranges in $x$, and the normalization condition (\ref{norm}), the Bernstein polynomials can be defined as a distribution. The error on $F^{n,k}_{2}$ is calculated from the error in $F_{2}^{exp}$ by using the Bernstein integrals, in quadrature, 
while the error on the $x$ values are obtained from the dispersion, 
\begin{equation}\label{eq:174}
(\Delta x)_{n,k}^{2} = \left| \left\langle x^2  \right\rangle_{n,k} - \left\langle x  \right\rangle_{n,k}^2 \right| = \frac{(n-k+1)}{(n+2)^{2} \, (n+3) } 
\end{equation}

We note that, because of their functional form, Eq.(\ref{eq:171}), every set of $n$ Bernstein moments can be written as a linear combination of Mellin moments.
By writing $B_{n,k}$ as a binomial expansion one has,
\begin{equation}
B_{n,k}(x) = \frac{(n+1)!}{k!} \sum_{l=0}^{n-k} \frac{(-1)^l}{l! (n-k-l)!} M_{l+k+1}
\label{binom}
\end{equation}
where $M_{l+k+1}$ are the Mellin moments,
\begin{equation}
M_{l+k+1} = \int\limits_0^1 F_2^{exp}(x, Q^2) \, x^{l+k} \, dx,
\end{equation}
namely, each Bernstein-averaged point corresponds to a specific combination of Mellin moments with coefficients determined by the binomial expansion (\ref{binom}). For example, for $n=2$, one obtains the following three equidistant points in $x$, 
\[ x_{20} = 0.25, \;\;\;  x_{21}=0.5, \;\;\; x_{22} =0.75, \]
and the weighted average of $F_2^{exp}$ is given by,
\begin{subequations}
\begin{eqnarray}
\label{berns3_x1}
F^{20}_2(x_{20},Q^2)& = & 3 M_{1} - 6 M_{2} + 3 M_{3},
\\
F^{21}_2(x_{21},Q^2) & = & 6 M_{2} - 6 M_3
,
\\   
F_2^{22}(x_{22},Q^2) & = & 3 M_3. 
\end{eqnarray}
\end{subequations}
One can see that how  the larger $l+k$ moments gradually contribute at larger $x$, while simultaneously  a {\em ``mathematically organized" mixing occurs at intermediate values of $x$ that simulates the way information is swapped in the experimental data}.

\begin{figure}
\begin{center}
\includegraphics[width=8.cm]{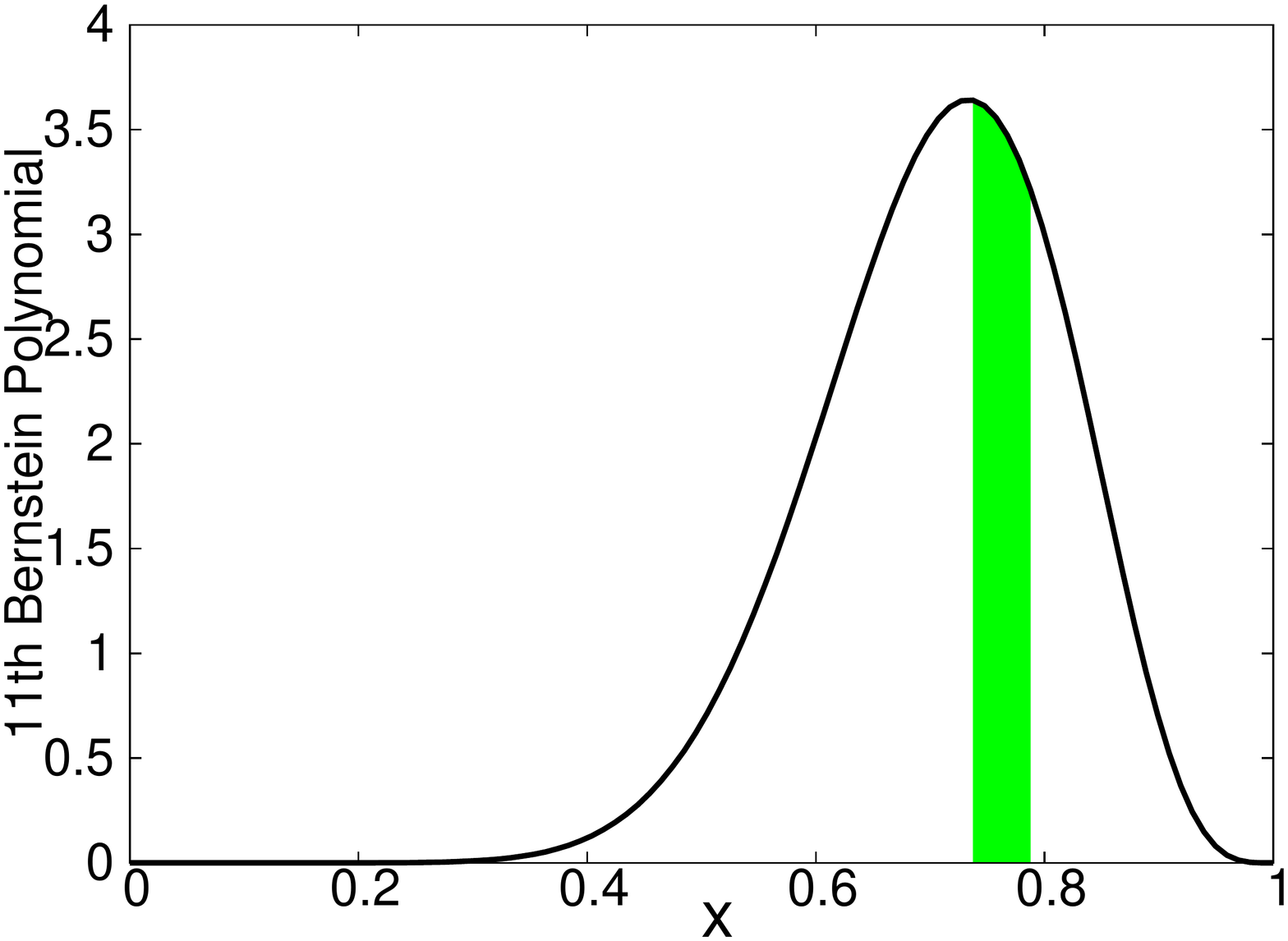}
\\
\includegraphics[width=8.cm]{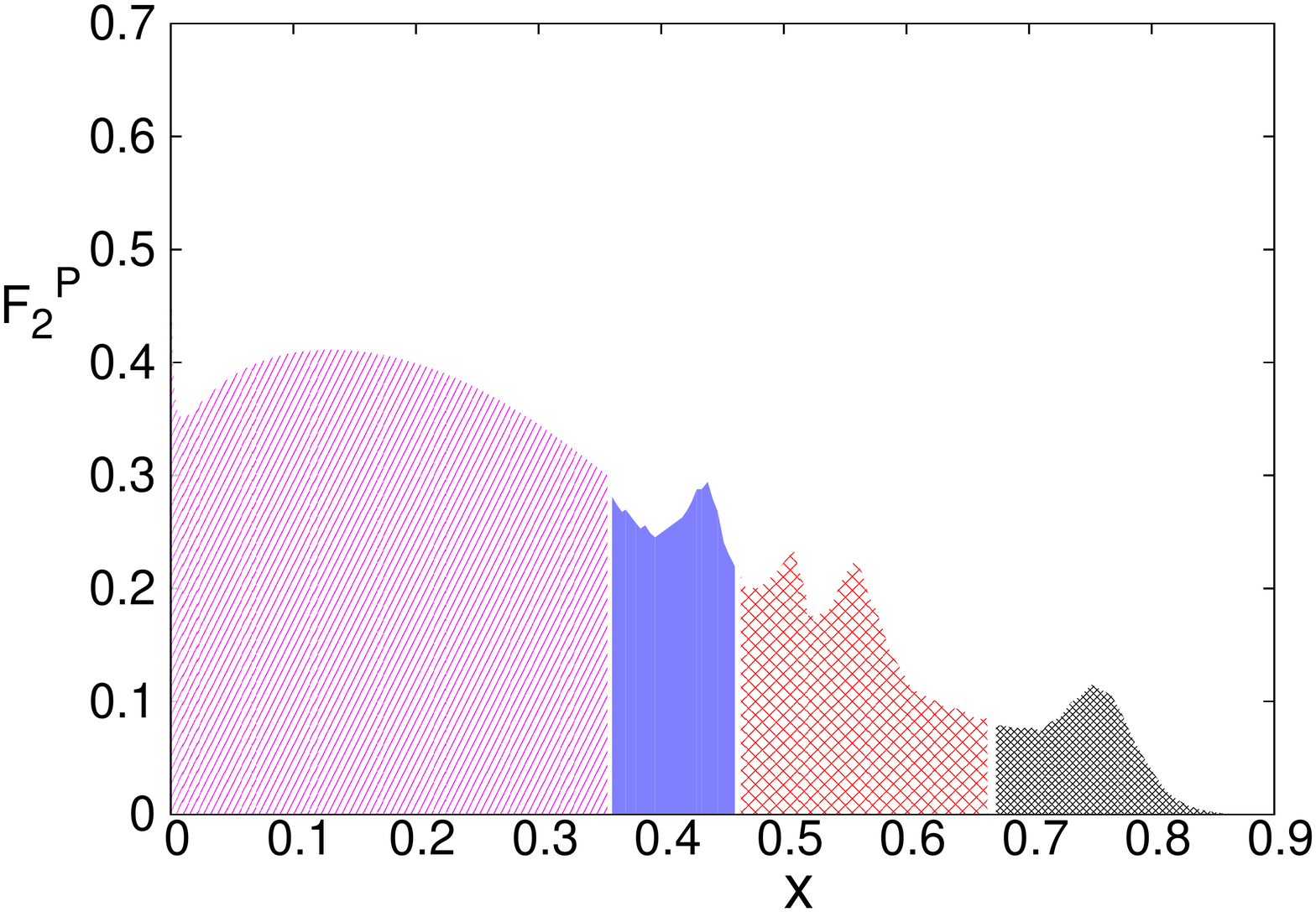}
\\
\includegraphics[width=8.cm]{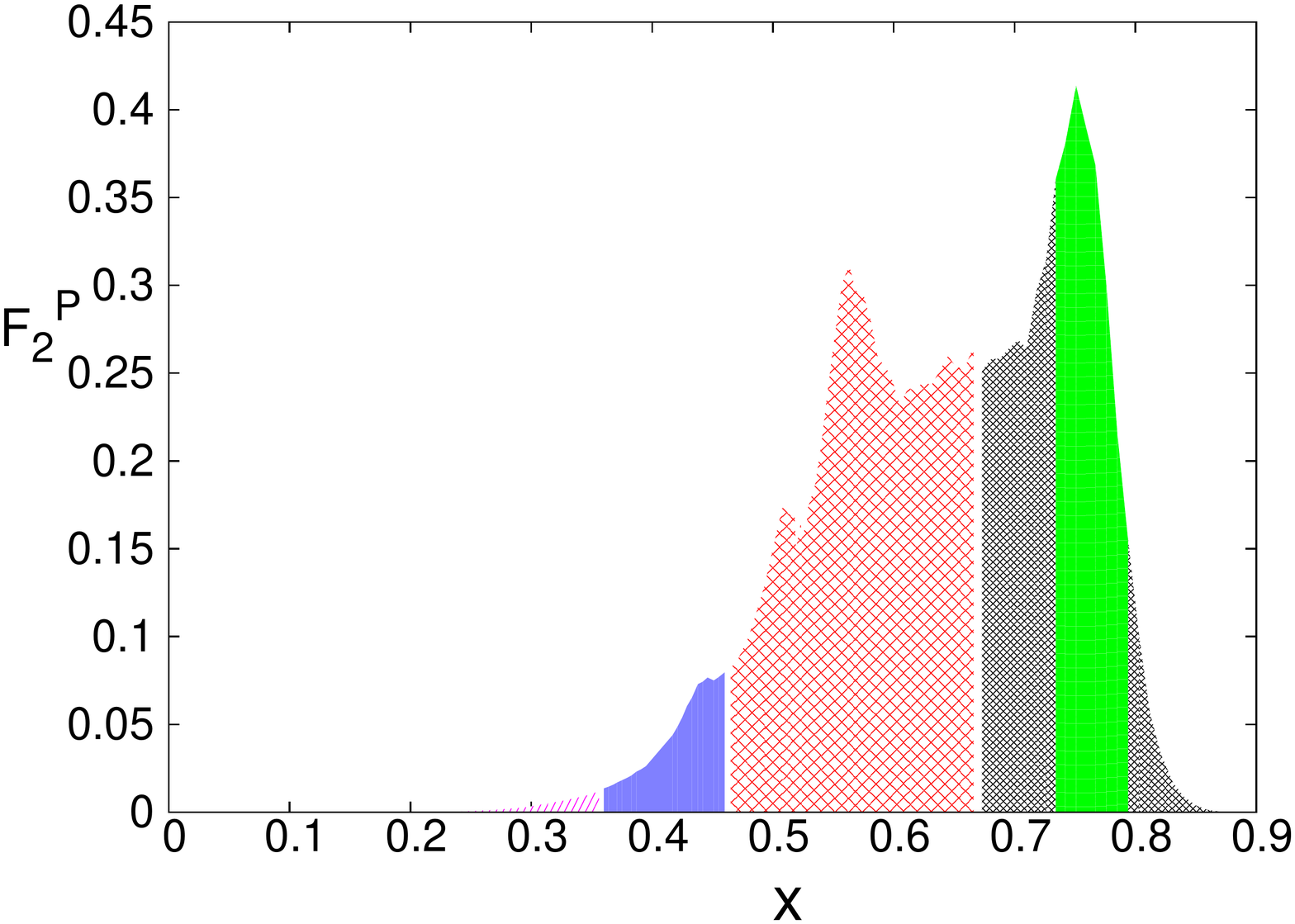}
\end{center}
\caption{(color online) $11^{th}$ Bernstein Polynomial is on the topmost plot and corresponding  $F_{2}^{P}$ Bernstein integrand for the $11^{th}$ moment  at $Q^{2} = 1.8$ GeV$^{2}$ is on the bottommost plot.  The shaded green stripe in both plots represent the region $\langle x \rangle \pm \Delta x$ where $\langle x \rangle$ is the computed Bernstein x value for the $11^{th}$ moment (Eq.(\ref{eq:173}) and $\Delta x$ is its error (Eq.(\ref{eq:174}).  In the middle plot, the entire spectrum in $x$ is shown divided into the four major kinematical regions (three major resonance regions, and the DIS region foremost to the left, for $W^2 \geq 4$ GeV$^2$, as described in the text).}
\label{fig:BernErrorRegion}
\end{figure}
In order to illustrate the working of the averaging procedure, in Figure \ref{fig:BernErrorRegion}
we show the integrand entering Eq.(\ref{eq:175}) and the corresponding average $x$ values for the kinematic bin at $Q^{2} = 1.8$ GeV$^{2}$.  
For our computations, $16$ Bernstein moments have been used so $n$ is set at $15$ and $k$ ranges from $0$ to $15$; in this case, $k=10$ as an example, would be used to determine the 11$^{th}$ Bernstein integral moment in the computation of $x$ in Eq.(\ref{eq:173}) and the 11$^{th}$ Bernstein integral moment of the error on $x$ in Eq.(\ref{eq:174}). 
In the chosen kinematic bin the resonance data clearly exhibit their structure with three prominent resonance regions: a first region dominated by $\Delta, P_{33}(1232)$, a second region dominated by the resonances $S_{11}(1535)$ 
and $D_{13}(1520)$, and a third region, dominated by $F_{15}(1680)$. 
For $W^2 > 4$ GeV$^2$ we are in the DIS region for which we used the structure functions from Ref.\cite{Lai:2010vv}. 
From the figure it is strikingly clear how the Bernstein polynomial suppresses the large $W^2$ regions. The average value of $F_2$ is dominated by the Delta region, however, the second resonance region also contributes with a lower probabilistic coefficient.

\begin{figure}
\includegraphics[width=8.cm]{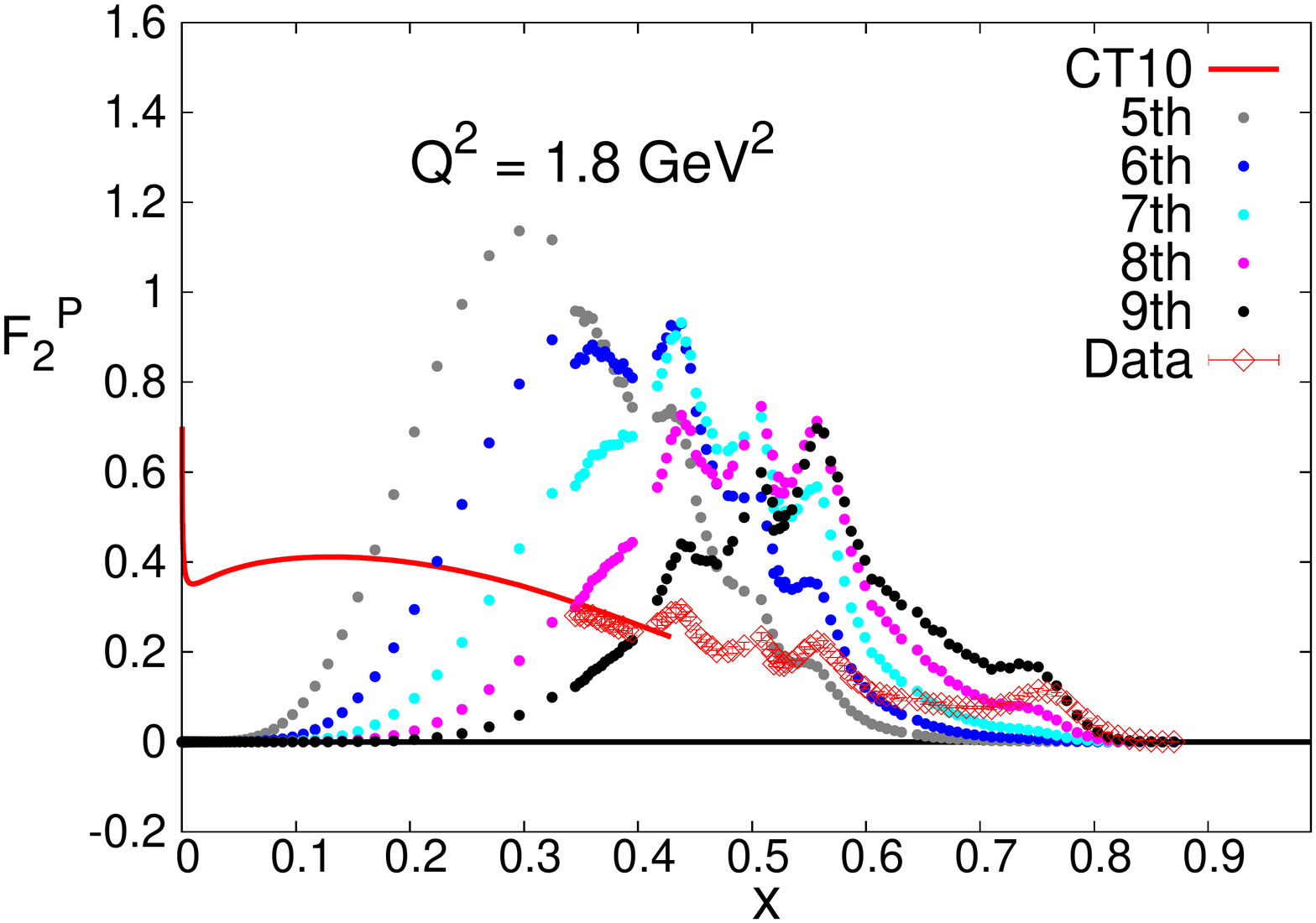}
\includegraphics[width=8.cm]{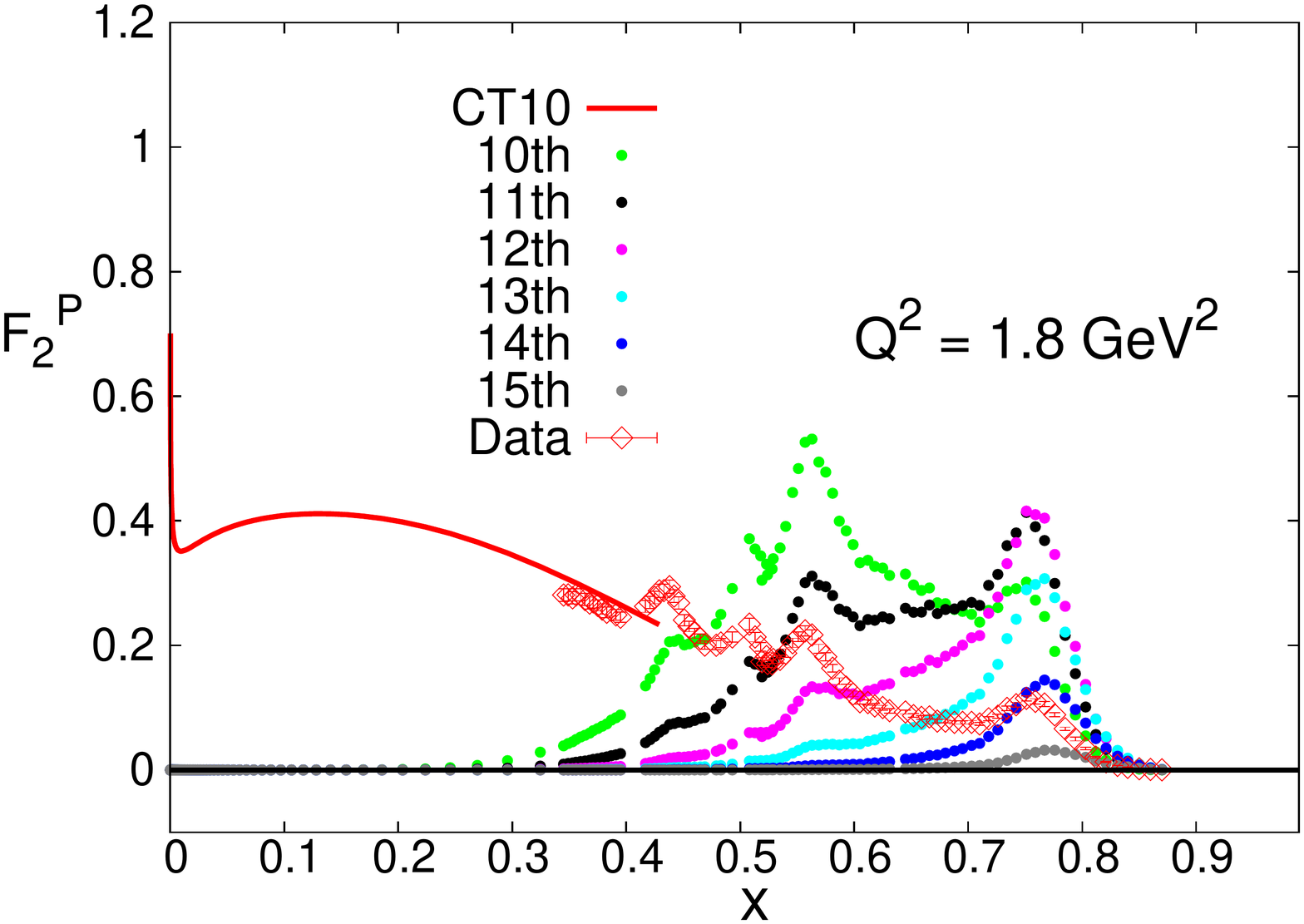}
\includegraphics[width=8.cm]{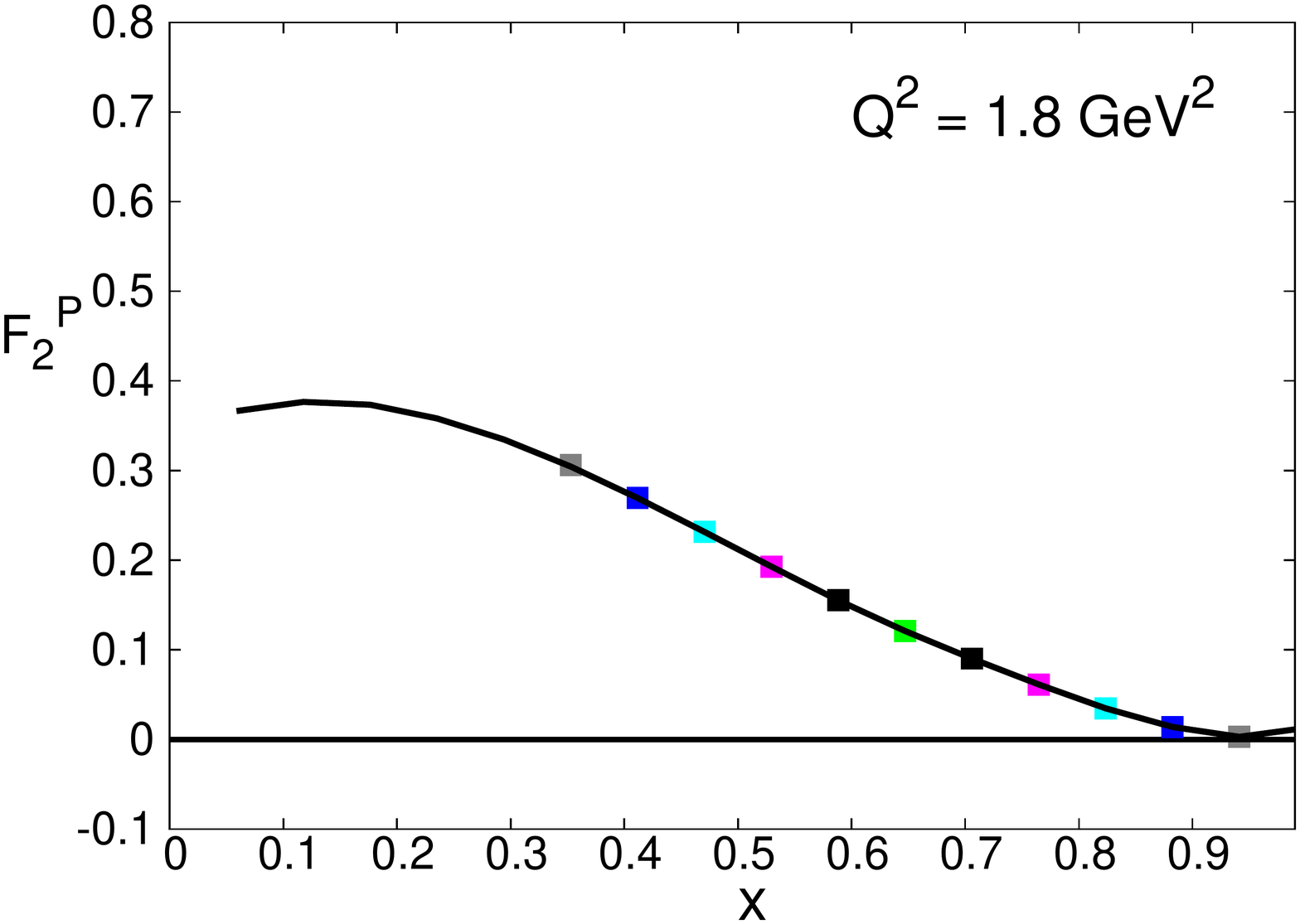}
\caption{(color online) $F_{2}^{P}$ resonance moment integrands and corresponding Bernstein moment points for $Q^{2} = 1.8$ GeV$^{2}$.  The $F_{2}^{P}$ Bernstein integrands, represented by the series of colored dots, are shown for the $5^{th}$,$6^{th}$, $7^{th}$, $8^{th}$ and $9^{th}$ moments in the topmost plot and for the $10^{th}$, $11^{th}$,$12^{th}$,$13^{th}$, $14^{th}$ and $15^{th}$ moments, in the central plot. The data points in the top two plots are a sample of available Jefferson Lab data used in our analysis for each kinematics \cite{hallc}. Each of these plots contains a continuation into the DIS region computed using the CT10 PDF parametrization from Ref.\cite{Lai:2010vv}.  The lower central figure's filled squares are the corresponding Bernstein moment points/averages for the integrands in the two upper plots. }
\label{fig:BernPolynomial2514}
\end{figure}
In Figure \ref{fig:BernPolynomial2514}, for the same kinematic bin ($Q^2= 1.8$ GeV$^2$) and using the same number of Bernstein moments ($n=16$) we show the integrands for various values of $0\leq k \leq n$ superimposed on the experimental data.  On the topmost panel we show the integrands for lower $k$ moments with $k=6,7,8,9$; on the central panel the corresponding integrands for the higher moments, with $k$ ranging from 10 to 15, are displayed. The filled squares are the integrated values, Eq.(\ref{eq:175}), calculated at the average $x$ according to Eq.(\ref{eq:173}). In order to calculate the moments we also need to evaluate the structure function outside the resonance region which we obtained from the CT10 global fit parametrized form in Ref.\cite{Lai:2010vv}. 

One can see how in the Bernstein integrals all resonances contribute with varying weights which are such that the lower moments, $k=6,7,8,9$, for our kinematics choice, are centered at $x \lesssim 0.65$, simultaneously emphasizing the higher mass resonances in the second and third region, while suppressing the contribution of the $\Delta$ resonance. As $k$ increases (right panel), and at larger values of $x$, all resonances contribute with similar weighting factors, until at the largest values of $x \approx 0.9$, the $\Delta$ region becomes dominant.     
The mixing of the various Mellin moments, which causes the resonance weighting, and the underlying probabilistic interpretation associated to the Bernstein polynomials distribution is what distinguishes the present averaging procedure from either local averaging, or truncated moments averaging. 

The line connecting the moments in Fig.\ref{fig:BernPolynomial2514} is a polynomial fit describing smooth curve that we later on use in our PDF fitting procedure. 

\begin{figure}[htp] \centering{
\includegraphics[width=8cm]{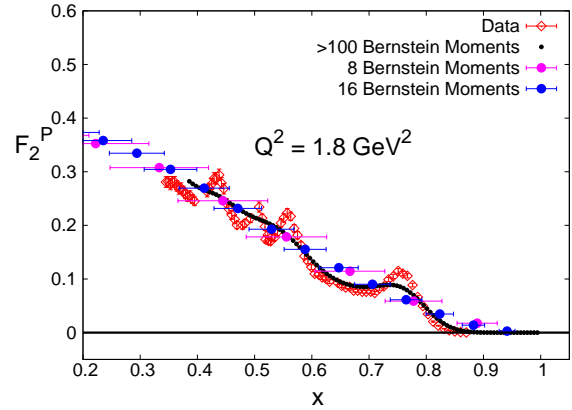}}
\caption[The $F_{2}^{P}$ values for the resonance region and Bernstein moments for $8$, $16$,$169$ resonance points]{The $F_{2}^{P}$ values for the resonance region and the choice of $16$ Bernstein moments, for large $x$ data calculations, are shown here along with a computation of $8$ Bernstein resonance points and a computation of $> 100$ to show how the Bernstein functions behave for larger numbers of chosen resonance points. The $Q^{2}$ range for the resonance data used for this figure is $[1.5:2.2]$.  $Q^{2} = 1.8 GeV^{2}$ is the average $Q^{2}$ value for the resonance data in this region. The red data points are a sample of available Jefferson Lab data used in our analysis for each kinematics \cite{hallc} for energy $E = 4$ GeV.  }
\label{fig:FinalBern34PPPMAX}
\end{figure}
Is there an ideal number of Bernstein polynomials to reproduce the average of $F_2$ in the resonance region, and does the averaging vary with this number?
This question is addressed in Fig.\ref{fig:FinalBern34PPPMAX} where we show three different evaluations of the average, for $n=7,15$ and $n>100$, including their dispersion in $x$. 
The reason for using $16$ Bernstein points is because the number of Bernstein moments needs to be large enough to capture the resonance behavior of the large $x$ data but not too large as to cause risk of over-fitting the data and/or replicating the behavior of the resonance data.  With $16$ data points the Bernstein averages of the resonance peaks for all the kinematics used in PDF fitting are sufficiently included in the large $x$ data computation and there is no risk of using an excessive number of data points.

From our study we conclude that while we cannot define a precise criterion to determine the ideal value of Bernstein polynomials, the best approach is to find a compromise between a minimum $n$ that gives a small enough $x$ dispersion, so that a smooth curve can be drawn, and a maximum $n$ which is small enough in order to avoid reproducing the resonance structure of the data (see the black dotted curve in Fig.\ref{fig:FinalBern34PPPMAX}). An additional guiding criterion is that the number of moments should not exceed the number of parameters of empirical resonance data fits, for instance the one in Ref.\cite{Christy:2007ve} which is 75. 
\footnote{Note that in Ref.\cite{Christy:2007ve} the fit includes also the structure function for longitudinal virtual photon polarization, $F_L$.}
Finally, we also note that by varying the number of Bernstein moments from 15 to 70 changes produces variations in the average curve by a few percent: this feature is already visible in Fig.\ref{fig:FinalBern34PPPMAX}, where it is shown even going from $n=7$ to 15 produces similar central values of the average. The difference between the two sets of moments is mostly in the uncertainty in $x$.  

\section{Results}
\label{sec3}
\vspace{-0.5cm}
\begin{figure}[htp]
\vspace{-0.5cm}
\includegraphics[width=7cm]{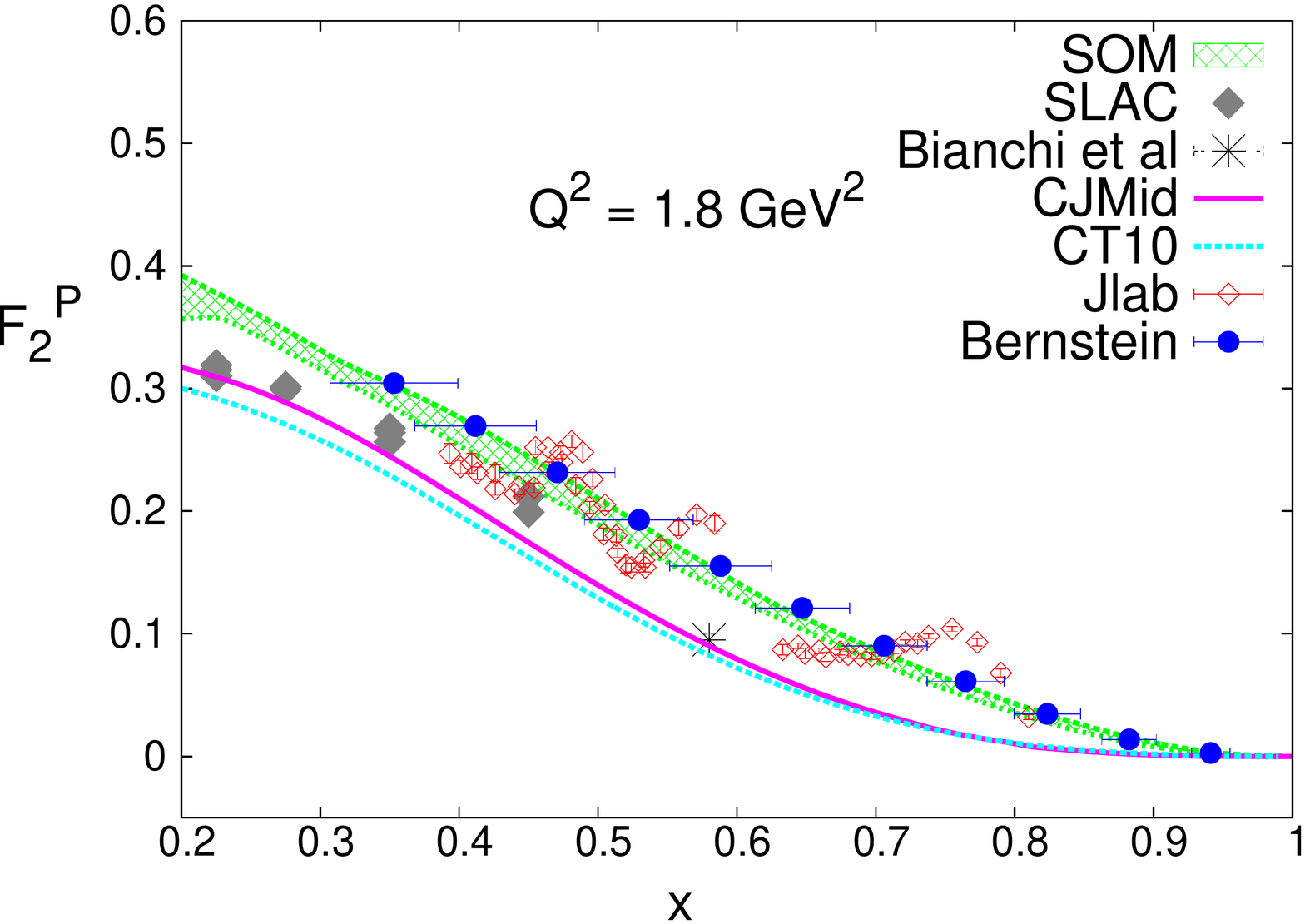}
\vspace{-0.5cm}
\includegraphics[width=7cm]{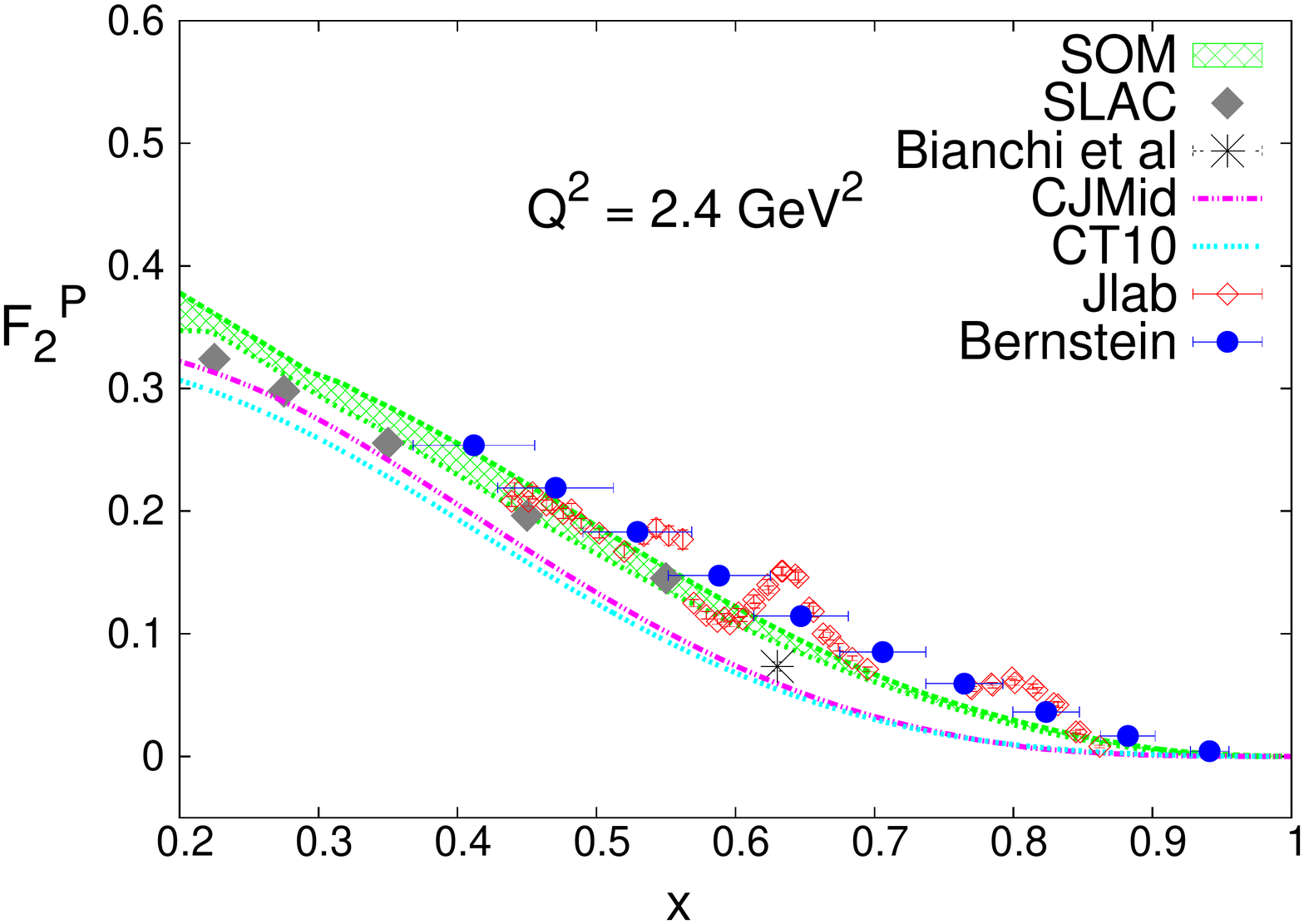}
\vspace{-0.5cm}
\includegraphics[width=7cm]{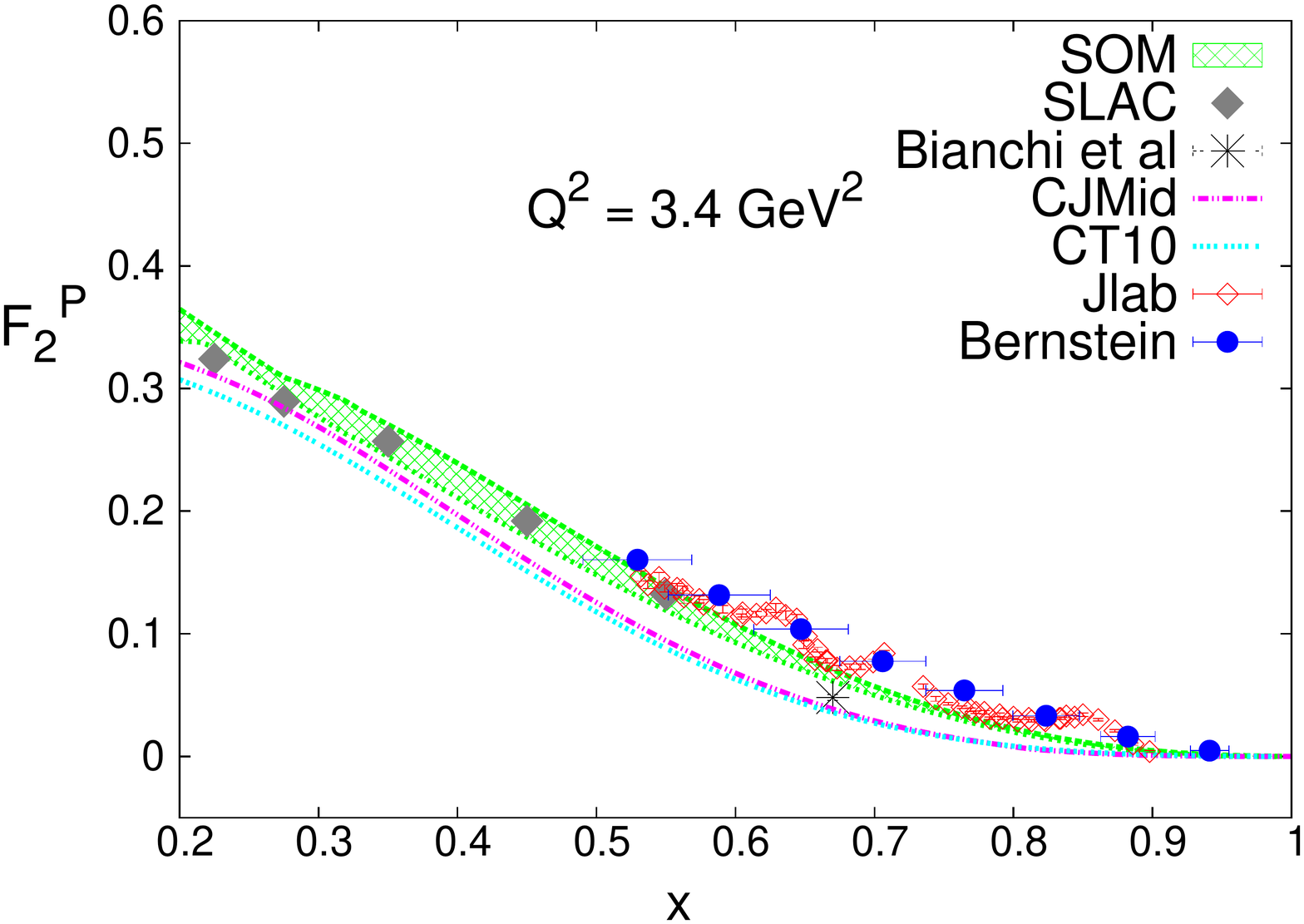}
\includegraphics[width=7cm]{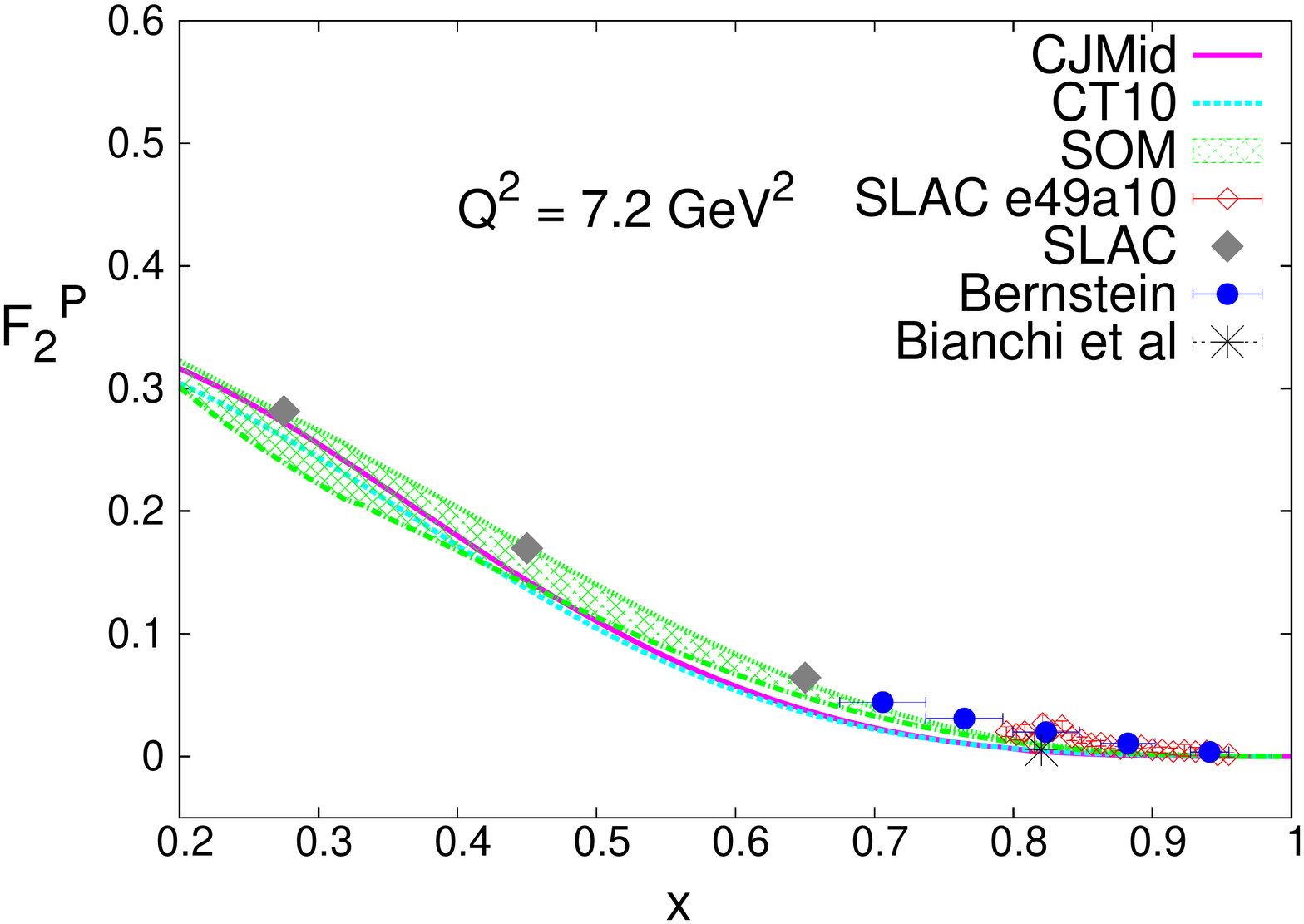}

\caption{(color online) The proton structure function, $F_{2}^{P}$, values 
for average $Q^{2}=$
1.8, 2.4, 3.4, and 7.2 GeV$^2$, plotted vs. $x$. The Bernstein points are fitted to experimental resonance data \cite{hallc,Stein:1975,Dasu:1988,Rock:1992,Bodek:1979,Stuart:1998,Dasu2:1988}. Shown are data sets representative of the resonance region \cite{Christy}. The diamonds are SLAC non resonant data \cite{Whitlow:1991uw}. The Bernstein moments are given by the blue points. For comparison we show PDF fits results from global fits \cite{Accardi:2011,Buckley:2014ana,Lai:2010vv}, and from the Self Organizing Map (SOM) generated structure functions \cite{Askanazi:2014gxa}.  Results from a previous $x$-averaging determination are given by the black point in each panel \cite{Bianchi:2003hi}. The $Q^2$ ranges  for the Jlab  data  the  are: $[1.7:1.9]$, $[2.3:2.5]$ and $[3.3:3.5]$ corresponding to average $Q^2$ values of 1.8, 2.5, 3.4 GeV$^2$. 
}
\label{fig:FinalBern18P}
\end{figure}

\begin{figure}[htp] 
\vspace{-0.5cm}
\includegraphics[width=7cm]{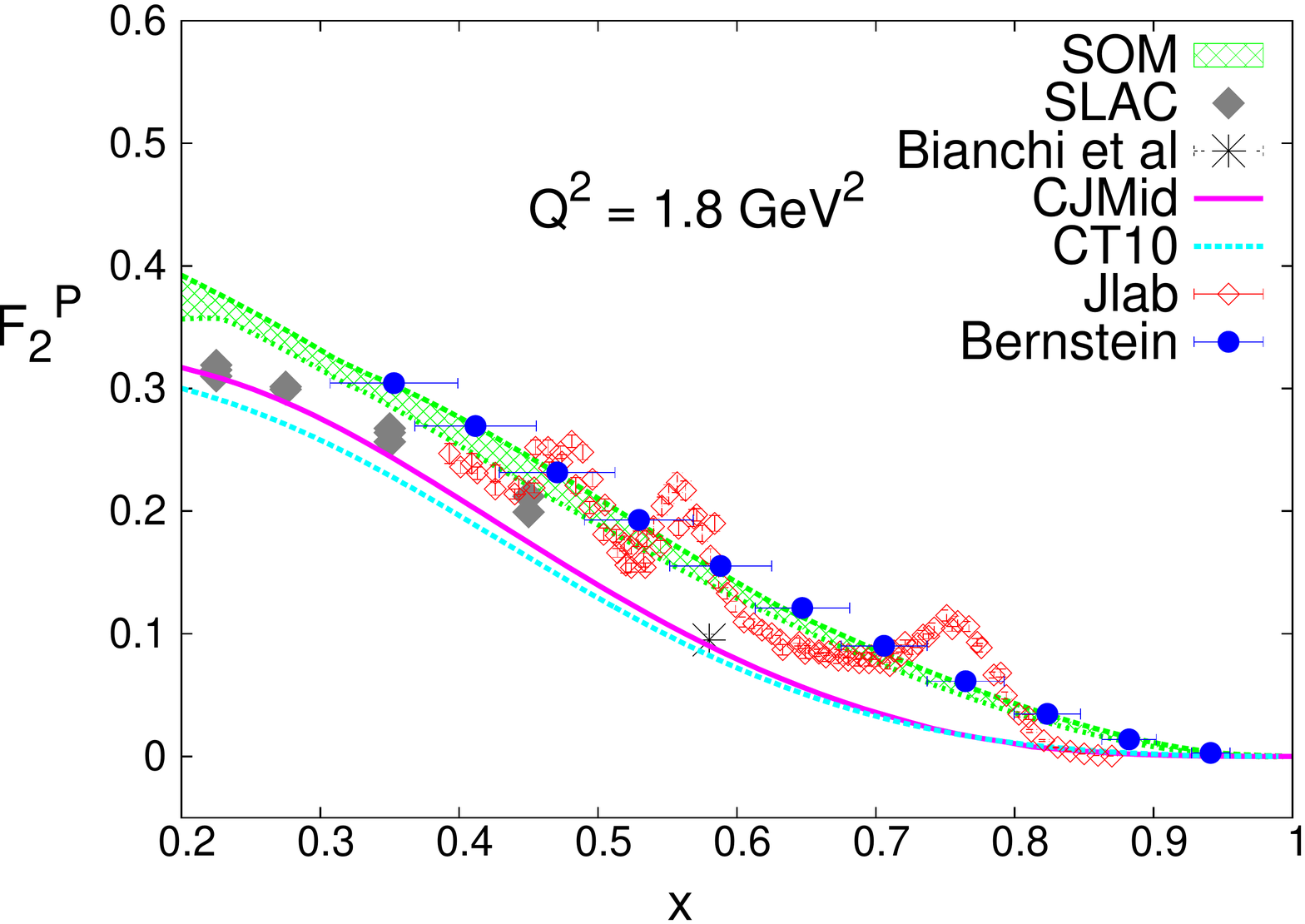}
\vspace{-0.5cm}
\includegraphics[width=7cm]{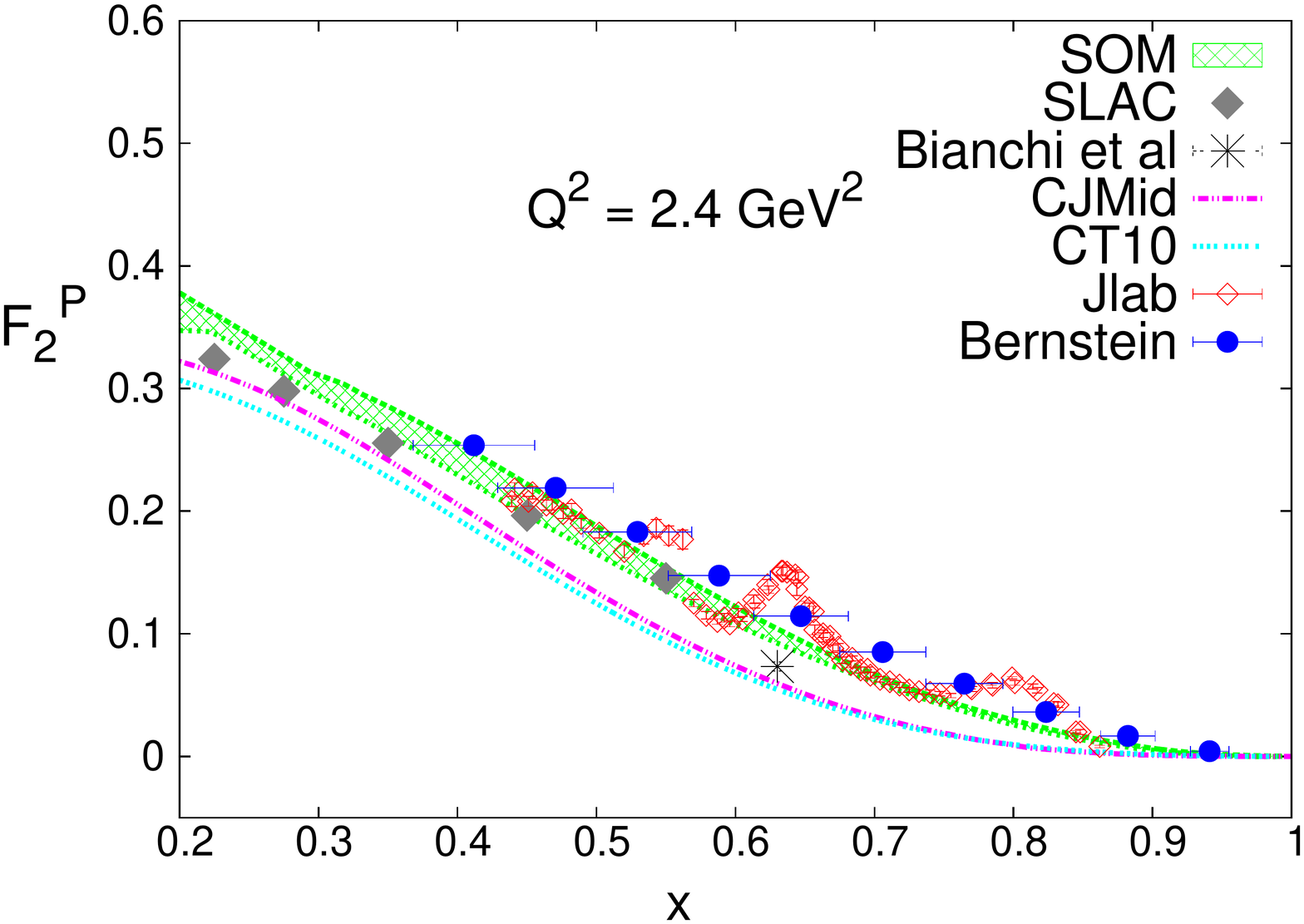}
\vspace{-0.5cm}
\includegraphics[width=7cm]
{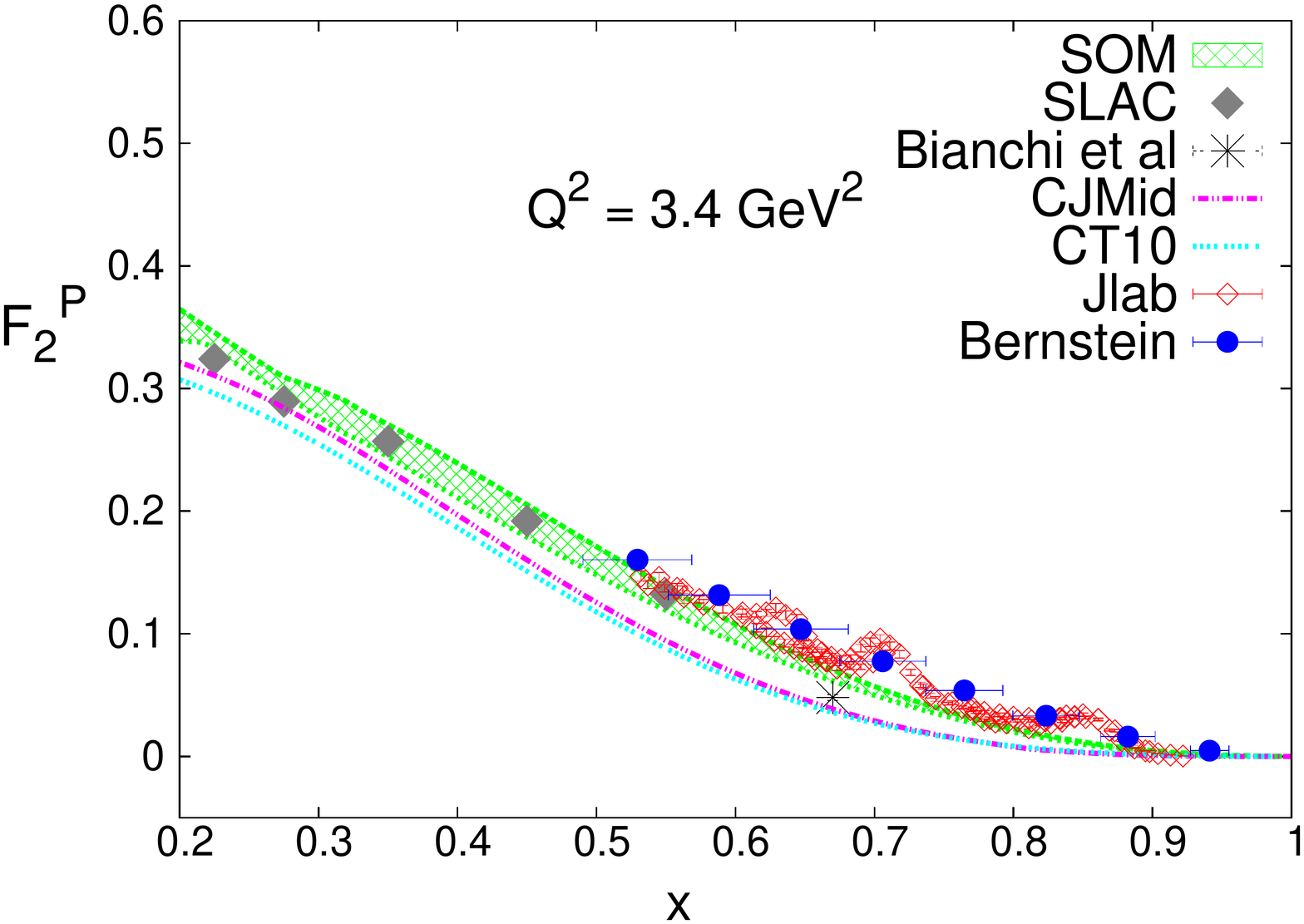}

\vspace{0.4cm}
\caption{(color online) The proton structure function, $F_{2}^{P}$, values for $Q^{2}=$
1.8, 2.4, 3.4, and 7.2 Gev$^2$, plotted vs. $x$. Notations for the Bernstein (blue points), PDFs \cite{Accardi:2011} \cite{Buckley:2014ana} (magenta) \cite{Lai:2010vv} (turquoise)  SOM  \cite{Askanazi:2014gxa} (green curves) SLAC \cite{Whitlow:1991uw} (diamonds) Bianchi \cite{Bianchi:2003hi} (black cross)  are the same as in Fig.\ref{fig:FinalBern18P}.  The experimental resonance data  \cite{Niculescu:2005rh} is also plotted. For the displayed Jlab Spectral data taken from  \cite{Christy}, the $Q^{2}$ ranges are $[1.7:1.9]$, $[2.3:2.5]$ and $[3.3:3.5]$ for the plots for  $Q^2$ values of 1.8, 2.5, 3.4 GeV$^2$. }
\label{fig:FinalBern18DW}
\end{figure}
The Bernstein-averaged structure functions along with the extrapolated curves  are shown in Figure \ref{fig:FinalBern18P} for the proton, and in Figure \ref{fig:FinalBern18DW} for the deuteron.   
The Bernstein points are fitted to experimental resonance data \cite{hallc} \cite{Stein:1975} \cite{Dasu:1988} \cite{Rock:1992} \cite{Bodek:1979} \cite{Stuart:1998} \cite{Dasu2:1988}. Our results are plotted against data sets representative of the resonance region for the $Q^{2}$ values of 1.8, 2.5, 3.4 GeV$^2$ \cite{Niculescu:2005rh} and 7 GeV$^2$ \cite{J.S.Poucher:1974}.  The diamonds are from the SLAC dataset including non resonant data \cite{Whitlow:1991uw}. For each kinematics we also show the result obtained from taking the average of $F_2$ over the whole resonance region (black square) \cite{Bianchi:2003hi}. 

\begin{table}[h]
\begin{center}
\begin{tabular}{|c|c|c|c|c|}
\hline
\hline
 $Q^{2} $ GeV$^{2}$ & Resonance    &  Bernstein Fit     &    Bianchi et al. \cite{Bianchi:2003hi} \\ 
\hline
\hline
  
   0.55    &        0.09728(42)        &    0.09735(26)      &   0.120      \\
   1.0       &      0.11494(27)        &    0.12111(20)    &     0.112        \\
   1.8       &      0.07418(20)        &    0.07106(12)    &     0.0905      \\
   2.5        &     0.05025(22)        &    0.05543(15)   &      0.0698         \\
   3.4          &   0.02796(12)        &    0.02911(8)      &    0.0402   \\
   5.7         &    0.00854(22)        &    0.00839(7)     &     0.00802    \\
   7.0           &  0.00453(20)        &    0.00522(8)      &    0.00531      \\
   8.2          &   0.00341(32)        &    0.00323(4)     &     0.00363     \\
   9.6         &    0.00184(29)        &    0.00215(3)    &         -   \\
      \hline
\hline
\end{tabular}
\caption{Integral values of resonance data points with errors, Bernstein moments with errors and functional forms for various $Q^{2}$ values.}
\label{table:funcint} 
\end{center}
 \end{table}
The theoretical curves in the figure are from the global QCD fits CT10 (NLO) \cite{Lai:2010vv} and CJMid \cite{Accardi:2011}, and from the Self-Organizing Maps (SOM) fitting procedure described below, which uses only inclusive electron proton and electron deuteron scattering data. 
The CT10 parametrization includes experimental
data from DIS, $W$ boson production, and single-inclusive jet production. In particular, including data from the Tevatron Run-II measurement of the forward-backward asymmetry in the rapidity distribution of the charged lepton in $W$ boson decay allows one to better constrain the ratio of the down to up quark PDFs, $d(x)/u(x)$.
The CJMid (CTEQ-Jefferson Lab) curve is from a a global fit of PDFs that  parametrizes a similar set of data including DIS, lepton pair creation, $W$ boson and jet production, focusing on the large $x$ region. 

The values of the Bernstein moments, $F_{2}(x_{k})$, corresponding to Figs.\ref{fig:FinalBern18P} and  \ref{fig:FinalBern18DW} are given in the Appendix in Tables \ref{table:18P} - \ref{table:7P}. Along with these values we show the separate contribution, in percentage, from the integrals in the $x$ range $[x_{min},x_{max}]$ where $x_{min}$ and $x_{max}$ are the minimum and maximum $x$ values of the resonance region in $x$ for each $Q^{2}$, Eqs.(\ref{eq:xminmax}), and from the integrals in the $x$ range $[0,x_{min}]$, calculated using the CT10 PDFs \cite{Lai:2010vv}.

For a quantitative check we evaluated also the integrals of the structure functions over the resonance region, computed directly from the data and from the fit to the Bernstein moments. Results are shown in Table \ref{table:funcint} compared with the results from Ref.\cite{Bianchi:2003hi}.  The integrations from the resonance data, and the Bernstein Moments fit show agreement with each other and in qualitative agreement with the previous extraction \cite{Bianchi:2003hi}.  

The Bernstein moments evaluations for the lowest $Q^2$ bin in Figs.\ref{fig:FinalBern18P} and  \ref{fig:FinalBern18DW} deviate sensibly from the data at lower values of $x$. The mismatching is more pronounced for the lower $Q^2$ bins. While on one side it is expected that Bernstein moments follow less closely the original curve at low $x$ \cite{Yndurain:1977wz}, the more pronounced deviations at low $Q^2$ might be interpreted as a consequence of the breaking up of perturbative QCD-based analyses at lower values of the scale.  

\begin{figure}
\includegraphics[width=8.cm]{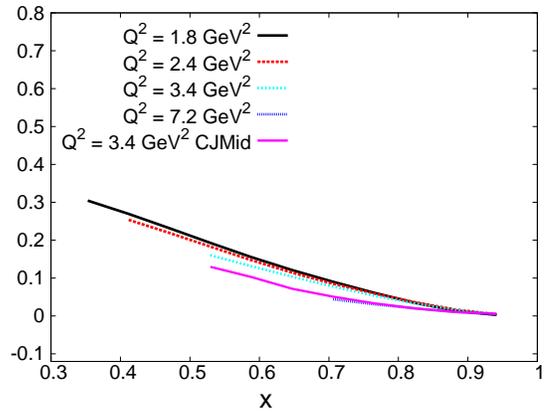}
\caption{(color online) The Bernstein smooth resonance curves for $Q^{2}= 1.8, 2.4, 3.4$ and $7.2$ {GeV}$^{2}$ plotted vs. $x$.  The CJMid curve with TMCs \cite{Accardi:2011} \cite{Buckley:2014ana} for $Q^{2} = 3.4$ {GeV}$^{2}$ is also plotted.}
\label{fig:BernPol}
\end{figure}
Figure \ref{fig:BernPol} 
illustrates the $Q^2$ dependence of our analysis for the largest $x$ bins. 
The theoretical curves are from the smooth curves from the fit (see Fig.\ref{fig:BernPolynomial2514} bottom panel) using the same $Q^{2}$ values as in Figs.\ref{fig:FinalBern18P},\ref{fig:FinalBern18DW}. Each curve is calculated in the corresponding resonance region range, $[x_{min}(Q^2), x_{max}(Q^2)]$.   The CJMid curve which includes TMCs \cite{Accardi:2011} for the $Q^{2}$ value of $3.4 {GeV}^{2}$ is shown for comparison. 

\begin{figure}[htp]
\includegraphics[width=8cm]{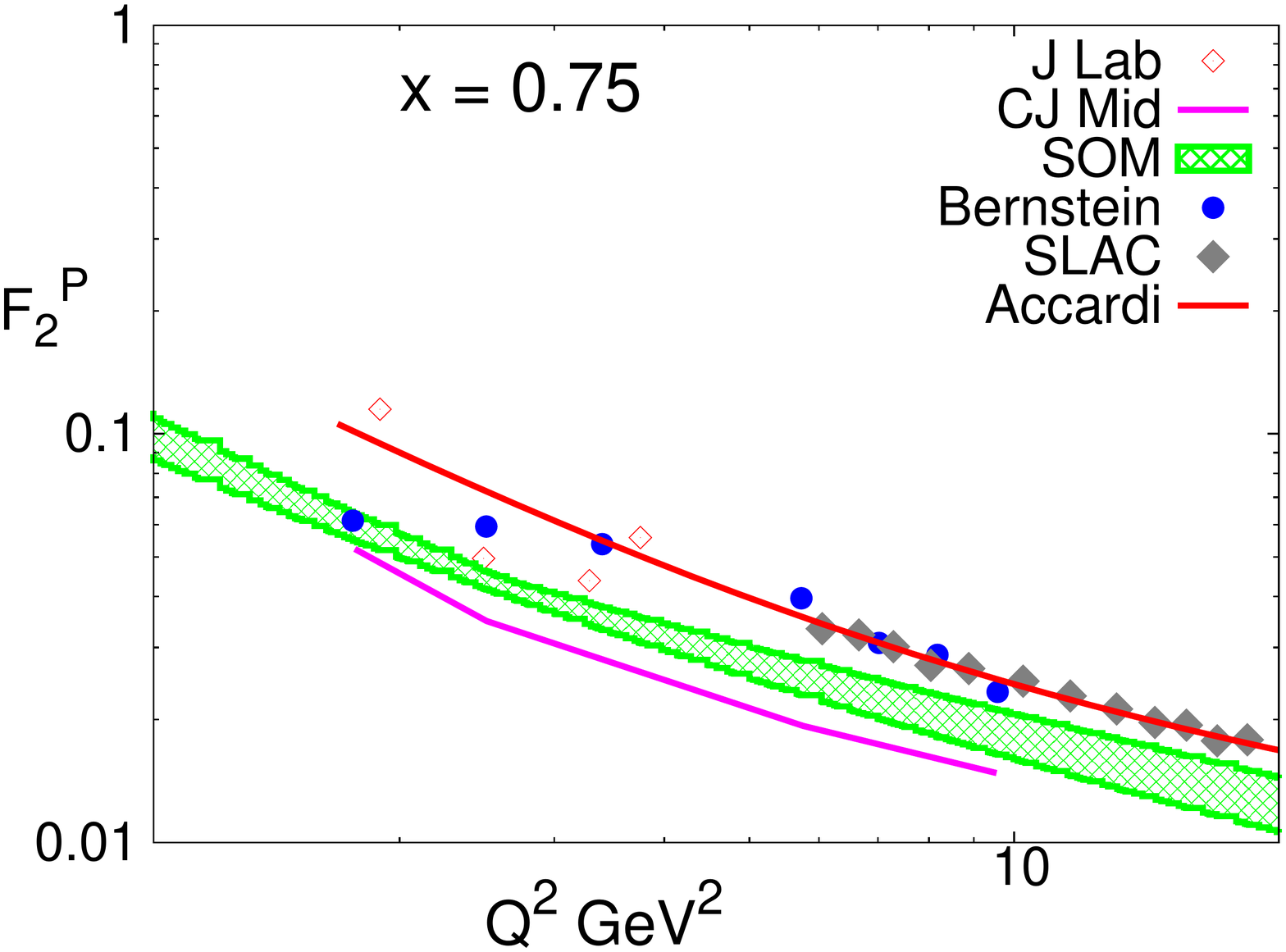}
\includegraphics[width=8cm]{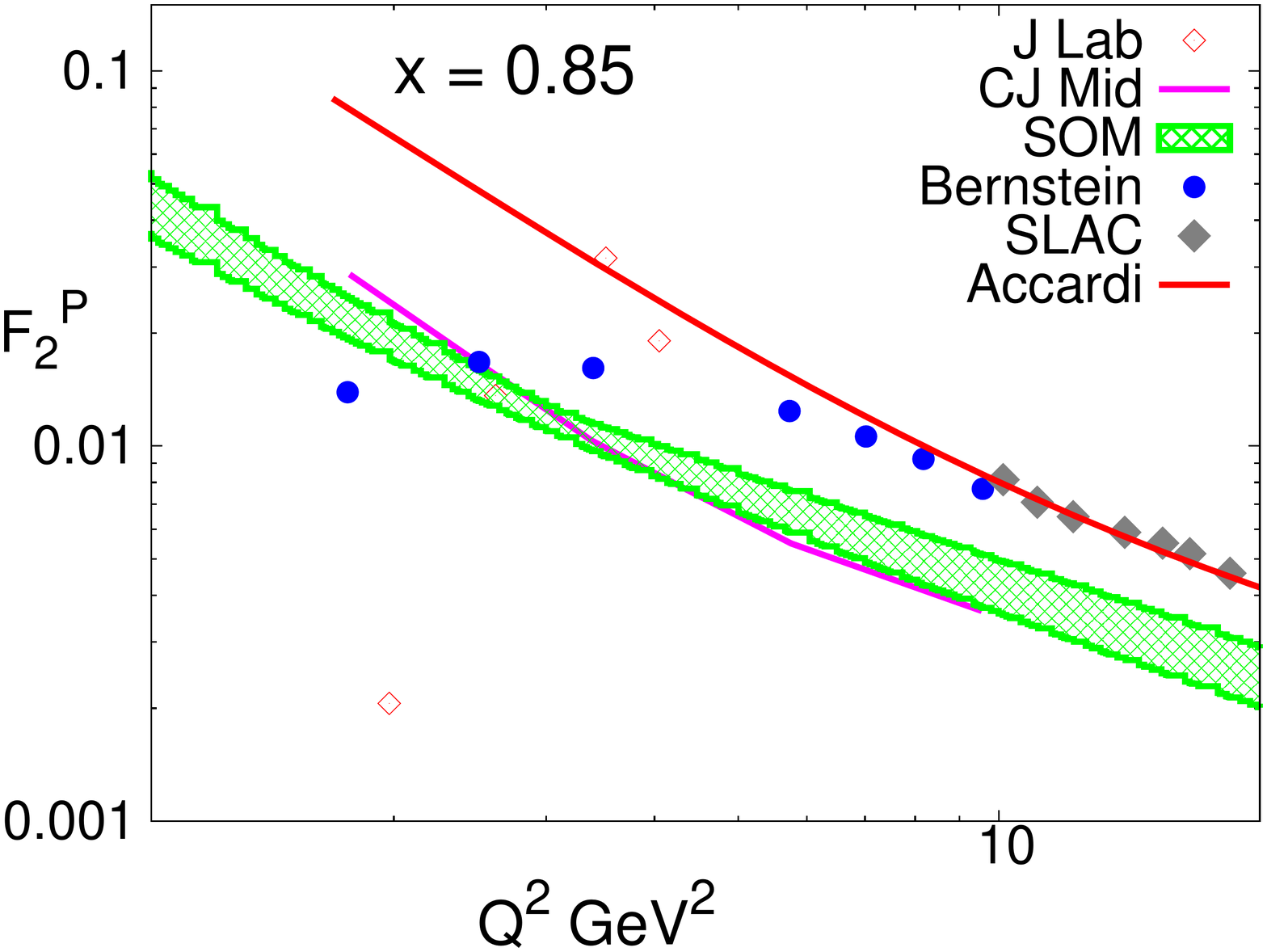}
\includegraphics[width=8cm]{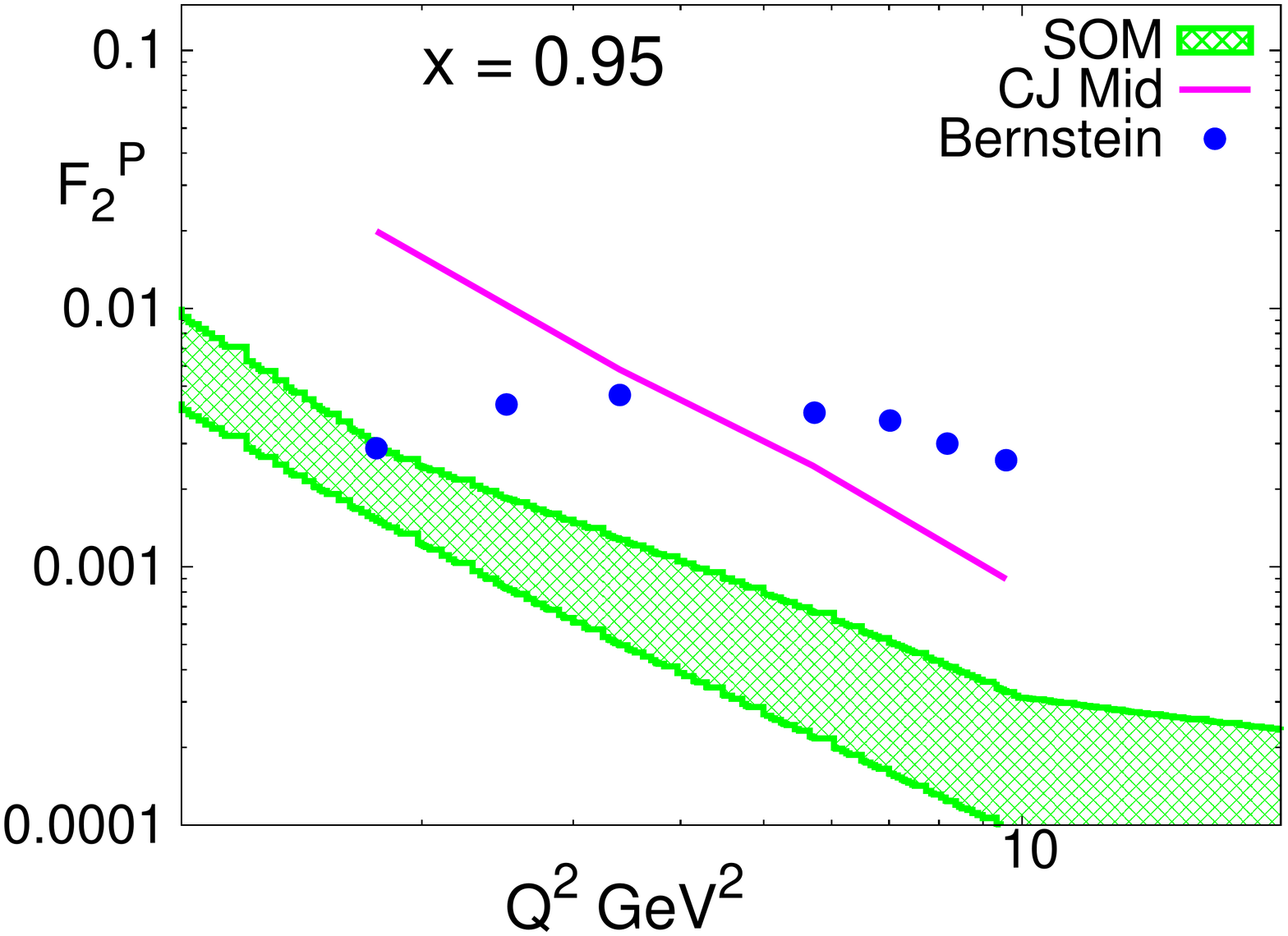}
\caption{(color online) $F_{2}^{P}$ values for the resonance region and the Bernstein moments  for $x = 0.75$ (top), $x = 0.85$ (center) and $x = 0.95$ (bottom).   The open red diamonds are the actual $F_2^P$ resonance data for the given $x$ value \cite{J.S.Poucher:1974,Malace:2009kw}; the blue points are the Bernstein averages for those points in the $Q^{2}$ range that constitutes the resonance region for the same $x$ value. The full diamonds are the available data in the DIS region \cite{J.S.Poucher:1974}. For comparison we show an extrapolation of the CJMid parametrization to the resonance region \cite{Accardi:2009br} \cite{Buckley:2014ana} (magenta), Accardi (red) \cite{Accardi:2016qay} and the SOM curve from the analysis in \cite{Askanazi:2014gxa} (green). 
}
\label{fig:FinalBern075xP}
\end{figure}
In Fig. \ref{fig:FinalBern075xP} we show  $F_{2}^{P}$ at large $x$ ($x = 0.75,0.85, 0.95$) plotted vs. $Q^2$, for $Q^2$ values ranging from the resonance region into the DIS region where experimental data from SLAC are available. The blue points at the lower $Q^2$ values are the Bernstein averages obtained from our analysis. One can see how these points blend in smoothly with the DIS experimental data points at the larger $Q^2$ values. One can also see a bend in the $Q^2$ behavior for low $Q^2$ values which clearly indicates the breakdown of PQCD-based analyses. This change in curvature happens at larger $Q^2$ values with increasing $x$.
The theoretical curves shown for comparison are from the CJMid parametrization including TMCs (magenta) \cite{Accardi:2009br}, the new fit from  Ref.\cite{Accardi:2016qay} (red) and the perturbative QCD contribution from the SOM analysis which includes both TMCs and Large $x$ resummation effects \cite{Askanazi:2014gxa}. While the curve from Ref.\cite{Accardi:2016qay} represents a good fit of the  $x=0.75$ and $x=0.85$ bins, the CJMid and SOM analyses miss the data, the most plausible explanation being that dynamical higher twist effects contribution is both present and substantial in this region in a wide range of $Q^2$.


\subsection{Self-Organizing Maps based fit}
Here we present an outline of the Self-Organizing-Maps (SOM) based code named SOMPDF \cite{Askanazi:2014gxa,Honkanen:2008mb} used in Figs.\ref{fig:FinalBern18P}, \ref{fig:FinalBern18DW}, \ref{fig:FinalBern075xP}. 

\begin{figure}[htp]
\includegraphics[width=8cm]{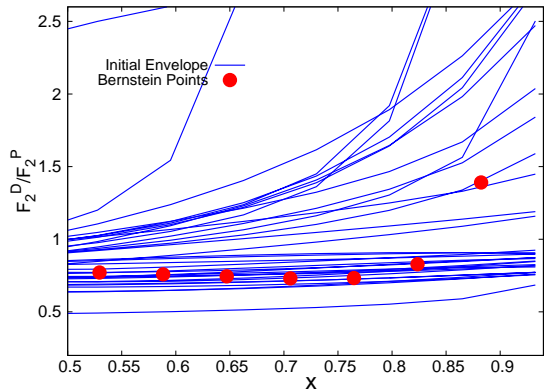}
\caption{An example of an envelope generated by the SOM is plotted for the ratio of the deuteron to the proton structure functions along with the same ratio for the Bernstein structure function points at $Q^{2} = 2.5$ GeV$^{2}$.   The envelope curves generated from the SOM algorithm encircle the collaboration data from above and below for an unbiased fit.  }
\label{fig:SOMEnvelope}
\end{figure}

A SOM is a specific type of Artificial Neural Network (ANN) \cite{Kohonen}. 
ANNs have been successfully used to parameterize the behavior of cross section components obtained from a variety of high energy physics experimental data:  the Neural Network PDFs obtained by the NNPDF collaboration are a primary example of this success \cite{Forte:2013wc}.  The NNPDF parametrization uses supervised learning which in turn utilizes the output data of the neural network as a set of reinforcements to train it.  With SOMPDFs we introduced a neural network based on {\it unsupervised learning} in which the PDFs are extracted by the network without the experimental data being used as a reinforcement in a continuous feedback mechanism.  In order to do this, we look to generate sets of PDFs by semi-randomly generating the parameters for a starting PDF set and selecting the PDFs for each subsequent iteration according to which ones have the best fit to experimental DIS and resonance experimental data. The PDFs generated from each iteration define an envelope that approaches the experimental data from the top and the bottom. An example of an envelope of PDFs is given in Fig.\ref{fig:SOMEnvelope} for the ratio $F_2^D/F_2^p$.  

The SOM is formed by a square map of cells, each one containing a variable number of PDFs from the envelope, which is used in the training step: each generated PDF is placed in a specific cell of this map.  An input vector is placed on the map that has the same dimension $n$ as each of the PDFs in the data cell.  The Euclidean distance of this input PDF to each PDF in the map is calculated, with the map PDF with the smallest distance being designated the Best Matching Unit (BMU).  The neighboring PDFs are adjusted by a neighborhood radius function inversely proportional to their distance from the BMU.  This process is called the activation, or training, of the map PDFs.  The radius of the neighborhood function becomes smaller with each iteration.  When a sufficient number of iterations is performed, the map PDFs form clusters based on similar properties of PDFs in specific sections of the map.  These clustering properties are particularly important in identifying specific trends and patterns of PDF data, for instance the PDFs' response to fitting the data when LxR, or nuclear effects are added.  This process is also advantageous because it enables us to visualize and organize multidimensional data on a two dimensional map.  SOMs are especially useful for fitting and analyzing data sets with the complexity and large number of dimensions that characterize the extraction of PDFs. More detailed information on both the fitting procedure and the error analysis can be found in Ref.\cite{Askanazi:2014gxa}. 
%

\section{Conclusions and Outlook}
\label{sec4}
In conclusion, we presented a novel approach for the analysis of  inclusive electron proton and electron deuteron scattering data in the resonance region based on the Bernstein polynomials technique. Bernstein polynomials are positive, orthogonal and they weigh the structure functions emphasizing its value in specific, calculable ranges in $x$. Moments of the structure function calculated using Bernstein polynomials can be written as linear combinations of Mellin moments, providing an easy connection with PQCD predictions. While the use of Bersntein Polynomials in DIS was already introduced at the inception of QCD as a method to reconstruct the structure function from its moments, we use it here in a different perspective which allows us to shed light on the averaging mechanism on which parton-hadron duality is founded.

Since we now dispose of a multitude of precise datasets in the resonance region, we can  
form a large number of moments, and study the averaging procedure as their number is increased gradually, and consequently the size of their ranges in $x$ is shortened. One can, in fact, define a critical number and interval size after which the smoothness of the curve is disrupted and the bumps of the resonance structure start reappearing. One therefore obtains quantitative clues for answering the question of how local the averaging phenomenon is. 

A dynamical picture emerges from the probabilistic characterization of our averaging procedure where, similarly to parton showers descriptions, the transition from the perturbative to the non-perturbative regime occurs through the formation of pre-hadronic substructures brought to light by the moments redistribution of strength. Future developments will include an extension using random Bernstein polynomials, for a detailed testing of the probabilistic interpretation behind duality. 

Besides providing new theoretical insight, this method is practically useful for mapping out the PDF large $x$ behavior in the multi-GeV region with smooth curves. Our interpolation results are stable if the number of Bernstein moments is $16 \lesssim n \lesssim 120$.  

To validate these ideas while simultaneously pinning down the structure of the $Q^2$ and $x$ dependence at very large $x$ including all sources of scaling violations and non-perturbative effects, will require an extended coverage of high precision data in both $x$ and $Q^2$. The Electron Ion Collider (EIC) will have such capability. 

\acknowledgements
We thank Donal Day and Nilanga Liyanage for support, Cynthia Keppel, Simona Malace and Wally Melnitchouk for many useful discussions, Eric Christy for sharing his views on the averaging procedure in the resonance region, and for providing us with his parametrization code, and Alberto Accardi for providing us with the most recent fit of the CJ analysis. We are grateful to Kathy Holcomb, Peter Alonzi, Bryan Wright and the UVACSE staff for their support with the U.Va. computing system. This work was completed under DOE grants DE-SC0016286 (S.L.) and DE-FG02-96ER40950 (E.A.).

\begin{widetext}
\appendix
\section{$F_2^P$ and $F_2^D$ Bernstein Moments Tables}
\label{app_a}
For completeness, we present the numerical values corresponding to Figures \ref{fig:FinalBern18P}, \ref{fig:FinalBern18DW}. Each table below corresponds to the $Q^2$ binning used in the figures. 

 \begin{table}[h]
\begin{center}
\caption{Bernstein moments for $n=16$ and for $Q^{2} = 1.8$ GeV$^{2}$. The moment number $k$ is displayed in column 1; their corresponding average $x$ values, in column 2, and the dispersion $\Delta x$, in column 3. The moments values are shown for the proton, $F_{n,k}^{(P)}$, in column 4, and for the deuteron, $F_{n,k}^{(D)}$, in column 8, their corresponding errors, $\Delta F_{n,k}^{(P)}$ and $\Delta F_{n,k}^{(D)}$ are shown in in column 5 and in column 9, respectively. The relative contributions of the resonance,  Res $F_{n,k}^{(P,D)}$ and DIS $DIS F_{n,k}^{(P,D)}$ regions are shown in columns 6 and 7 for the proton, and columns 10 and 11 for the deuteron.}
\begin{tabular}{|c|c|c|c|c|c|c|c|c|c|c|}
\hline
\hline
  $k$  & $x$   & $\Delta$ $x$   &  $F_{n,k}^{(P)}(Q^{2})$  & $\Delta F_{n,k}^{(err P )}(Q^{2})$ &  Res $F_{n,k}^{(P)}(Q^{2})$  & DIS     $F_{n,k}^{(P)}(Q^{2})$  &   $F_{n,k}^{(D)}(Q^{2})$  & $\Delta F_{n,k}^{(err D )}(Q^{2})$ &  Res $F_{n,k}^{(D)}(Q^{2})$  & DIS     $F_{n,k}^{(D)}(Q^{2})$        \\
\hline
\hline
   0                &        0.0588             &        0.0555             &     0.3664                 &     0.3892E-02             &        0.0009             &        0.9987             & 0.2772 & 0.3358E-02 & 0.0073 & 0.9927 \\
      1                 &        0.1176             &        0.0537             &     0.3770                 &     0.3763E-02             &        0.0078             &        0.9887             & 0.1727 & 0.2346E-02 & 0.0933 & 0.9067 \\
      2                 &        0.1765             &        0.0519             &     0.3745                 &     0.3519E-02             &        0.0348             &        0.9522             & 0.1248 & 0.1403E-02 & 0.4864 & 0.5136 \\
      3                 &        0.2353             &        0.0500             &     0.3606                 &     0.3188E-02             &        0.1024             &        0.8671             & 0.1626 & 0.1817E-02 & 0.8809 & 0.1191 \\
      4                 &        0.2941             &        0.0480             &     0.3376                 &     0.2793E-02             &        0.2239             &        0.7250             & 0.2427 & 0.2776E-02 & 0.9817 & 0.0183 \\
      5                 &        0.3529             &        0.0460             &     0.3062                 &     0.2239E-02             &        0.3914             &        0.5438             & 0.3000 & 0.3191E-02 & 0.9974 & 0.0026 \\
      6                 &        0.4118             &        0.0438             &     0.2682                 &     0.1573E-02             &        0.5765             &        0.3591             & 0.2992 & 0.2787E-02 & 0.9996 & 0.0004 \\
      7                &        0.4706             &        0.0416             &     0.2272                 &     0.9900E-03             &        0.7442             &        0.2049             & 0.2505 & 0.1895E-02 & 1.0000 & 0.0000 \\
      8                &        0.5294             &        0.0392             &     0.1871                 &     0.6280E-03             &        0.8687             &        0.0989             & 0.1866 & 0.1037E-02 & 1.0000 & 0.0000 \\
      9               &        0.5882             &        0.0367             &     0.1504                 &     0.4530E-03             &        0.9440             &        0.0396             & 0.1322 & 0.5120E-03 & 1.0000 & 0.0000 \\
      10                 &        0.6471             &        0.0340             &     0.1178                 &     0.3640E-03             &        0.9805             &        0.0130             & 0.9350E-01 & 0.2980E-03 & 1.0000 & 0.0000 \\
      11                &        0.7059             &        0.0310             &     0.8790E-01             &     0.3070E-03             &        0.9945             &        0.0034             & 0.6659E-01 & 0.2180E-03 & 1.0000 & 0.0000 \\
      12                 &        0.7647             &        0.0277             &     0.5918E-01             &     0.2490E-03             &        0.9987             &        0.0008             & 0.4642E-01 & 0.1650E-03 & 1.0000 & 0.0000 \\
      13               &        0.8235             &        0.0240             &     0.3273E-01             &     0.1700E-03             &        0.9998             &        0.0002             & 0.3132E-01 & 0.1270E-03 & 1.0000 & 0.0000 \\
      14                 &        0.8824             &        0.0196             &     0.1282E-01             &     0.8400E-04             &        1.0000             &        0.0000             & 0.2306E-01 & 0.1090E-03 & 1.0000 & 0.0000 \\
      15                &        0.9412             &        0.0139             &     0.2609E-02             &     0.2200E-04             &        1.0000             &        0.0000             & 0.2327E-01 & 0.1510E-03 & 1.0000 & 0.0000 \\

 \hline
\hline


\end{tabular}

\label{table:18P} 
\end{center}
 \end{table}

\begin{table}[h]
\begin{center}
\caption{Same as Table \ref{table:18P} for $Q^{2} = 2.5$ GeV$^{2}$.}
\begin{tabular}{|c|c|c|c|c|c|c|c|c|c|c|}
\hline
\hline
  $k$  & $x$   & $\Delta$ $x$   &  $F_{n,k}^{(P)}(Q^{2})$  & $\Delta F_{n,k}^{(err P )}(Q^{2})$ &  Res $F_{n,k}^{(P)}(Q^{2})$  & DIS     $F_{n,k}^{(P)}(Q^{2})$  &   $F_{n,k}^{(D)}(Q^{2})$  & $\Delta F_{n,k}^{(err D )}(Q^{2})$ &  Res $F_{n,k}^{(D)}(Q^{2})$  & DIS     $F_{n,k}^{(D)}(Q^{2})$     \\
\hline
\hline
     0                 &        0.0588             &        0.0555             &     0.3822                 &     0.3736E-02             &        0.0001             &        0.9998  &   0.3644                 &     0.3418E-02             &        0.0001             &        0.9999   \\
1                 &        0.1176             &        0.0537             &     0.3785                 &     0.3594E-02             &        0.0009             &        0.9981             & 0.3481 & 0.3315E-02 & 0.0008 & 0.9984 \\
      2                 &        0.1765             &        0.0519             &     0.3664                 &     0.3346E-02             &        0.0059             &        0.9891             & 0.3259 & 0.3188E-02 & 0.0052 & 0.9903 \\
      3                 &        0.2353             &        0.0500             &     0.3446                 &     0.3092E-02             &        0.0241             &        0.9586             & 0.2969 & 0.3018E-02 & 0.0218 & 0.9623 \\
      4                 &        0.2941             &        0.0480             &     0.3170                 &     0.2939E-02             &        0.0717             &        0.8866             & 0.2653 & 0.2843E-02 & 0.0664 & 0.8938 \\
      5                 &        0.3529             &        0.0460             &     0.2864                 &     0.2780E-02             &        0.1650             &        0.7596             & 0.2336 & 0.2577E-02 & 0.1560 & 0.7702 \\
      6                 &        0.4118             &        0.0438             &     0.2537                 &     0.2461E-02             &        0.3073             &        0.5874             & 0.2026 & 0.2144E-02 & 0.2952 & 0.5996 \\
     7                 &        0.4706             &        0.0416             &     0.2190                 &     0.1948E-02             &        0.4815             &        0.4021             & 0.1717 & 0.1578E-02 & 0.4680 & 0.4136 \\
      8                 &        0.5294             &        0.0392             &     0.1830                 &     0.1354E-02             &        0.6570             &        0.2396             & 0.1411 & 0.1010E-02 & 0.6443 & 0.2487 \\
      9                 &        0.5882             &        0.0367             &     0.1474                 &     0.8480E-03             &        0.8038             &        0.1221             & 0.1118 & 0.5780E-03 & 0.7940 & 0.1281 \\
      10                 &        0.6471             &        0.0340             &     0.1143                 &     0.5160E-03             &        0.9055             &        0.0520             & 0.8515E-01 & 0.3300E-03 & 0.8991 & 0.0553 \\
      11                 &        0.7059             &        0.0310             &     0.8508E-01             &     0.3280E-03             &        0.9627             &        0.0180             & 0.6215E-01 & 0.2080E-03 & 0.9594 & 0.0195 \\
      12                 &        0.7647             &        0.0277             &     0.5933E-01             &     0.2250E-03             &        0.9881             &        0.0050             & 0.4346E-01 & 0.1420E-03 & 0.9872 & 0.0054 \\
      13                 &        0.8235             &        0.0240             &     0.3621E-01             &     0.1640E-03             &        0.9970             &        0.0011             & 0.2998E-01 & 0.1000E-03 & 0.9971 & 0.0011 \\
      14                 &        0.8824             &        0.0196             &     0.1674E-01             &     0.1060E-03             &        0.9993             &        0.0002             & 0.2329E-01 & 0.8800E-04 & 0.9996 & 0.0001 \\
      15                 &        0.9412             &        0.0139             &     0.4250E-02             &     0.4100E-04             &        1.0000             &        0.0000             & 0.2247E-01 & 0.1260E-03 & 1.0000 & 0.0000 \\

 \hline
\hline


\end{tabular}

\label{table:25P} 
\end{center}
 \end{table}

\begin{table}[h]
\begin{center}
\caption{Same as Table \ref{table:18P} for  $Q^{2} = 3.4$ GeV$^{2}$.}
\begin{tabular}{|c|c|c|c|c|c|c|c|c|c|c|}
\hline
\hline
  $k$  & $x$   & $\Delta$ $x$   &  $F_{n,k}^{(P)}(Q^{2})$  & $\Delta F_{n,k}^{(err P )}(Q^{2})$ &  Res $F_{n,k}^{(P)}(Q^{2})$  & DIS     $F_{n,k}^{(P)}(Q^{2})$  &   $F_{n,k}^{(D)}(Q^{2})$  & $\Delta F_{n,k}^{(err D )}(Q^{2})$ &  Res $F_{n,k}^{(D)}(Q^{2})$  & DIS     $F_{n,k}^{(D)}(Q^{2})$     \\
\hline
\hline
    0                 &        0.0588             &        0.0555             &     0.3949                 &     0.3624E-02             &        0.0000             &        1.0000             & 0.3766 & 0.3316E-02 & 0.0000 & 1.0000 \\
      1                 &        0.1176             &        0.0537             &     0.3799                 &     0.3468E-02             &        0.0000             &        0.9999             & 0.3491 & 0.3201E-02 & 0.0000 & 1.0000 \\
      2                 &        0.1765             &        0.0519             &     0.3611                 &     0.3209E-02             &        0.0004             &        0.9995            & 0.3209 & 0.3061E-02 & 0.0003 & 0.9996 \\
      3                 &        0.2353             &        0.0500             &     0.3332                 &     0.2953E-02             &        0.0026             &        0.9970             & 0.2868 & 0.2888E-02 & 0.0021 & 0.9975 \\
      4                 &        0.2941             &        0.0480             &     0.2987                 &     0.2792E-02             &        0.0112             &        0.9875             & 0.2494 & 0.2722E-02 & 0.0092 & 0.9894 \\
      5                &        0.3529             &        0.0460             &     0.2613                 &     0.2640E-02             &        0.0369             &        0.9593             & 0.2119 & 0.2511E-02 & 0.0309 & 0.9649 \\
      6                 &        0.4118             &        0.0438             &     0.2247                 &     0.2397E-02             &        0.0969             &        0.8953             & 0.1769 & 0.2217E-02 & 0.0828 & 0.9081 \\
      7                 &        0.4706             &        0.0416             &     0.1911                 &     0.2041E-02             &        0.2060             &        0.7810             & 0.1457 & 0.1847E-02 & 0.1800 & 0.8044 \\
      8                 &        0.5294             &        0.0392             &     0.1604                 &     0.1591E-02             &        0.3627             &        0.6204             & 0.1180 & 0.1417E-02 & 0.3251 & 0.6539 \\
      9                 &        0.5882             &        0.0367             &     0.1315                 &     0.1101E-02             &        0.5434             &        0.4391             & 0.9314E-01 & 0.9650E-03 & 0.5009 & 0.4764 \\
      10                 &        0.6471             &        0.0340             &     0.1037                 &     0.6680E-03             &        0.7140             &        0.2715             & 0.7080E-01 & 0.5620E-03 & 0.6767 & 0.3036 \\
      11                 &        0.7059             &        0.0310             &     0.7750E-01             &     0.3750E-03             &        0.8472             &        0.1431             & 0.5135E-01 & 0.2810E-03 & 0.8224 & 0.1641 \\
      12                 &        0.7647             &        0.0277             &     0.5372E-01             &     0.2350E-03             &        0.9324             &        0.0626             & 0.3524E-01 & 0.1370E-03 & 0.9206 & 0.0722 \\
      13                 &        0.8235             &        0.0240             &     0.3305E-01             &     0.1830E-03             &        0.9758             &        0.0221             & 0.2256E-01 & 0.8800E-04 & 0.9727 & 0.0245 \\
      14                 &        0.8824             &        0.0196             &     0.1614E-01             &     0.1410E-03             &        0.9931             &        0.0063             & 0.1324E-01 & 0.8700E-04 & 0.9935 & 0.0057 \\
      15                &        0.9412             &        0.0139             &     0.4626E-02             &     0.6600E-04             &        0.9985             &        0.0015             & 0.7927E-02 & 0.1930E-03 & 0.9992 & 0.0006 \\

\hline
\hline


\end{tabular}

\label{table:34P} 
\end{center}
 \end{table}


\begin{table}[h]
\begin{center}
\caption{Same as Table \ref{table:18P} for $Q^{2} = 7$ GeV$^{2}$.}
\begin{tabular}{|c|c|c|c|c|c|c|c|c|c|c|}
\hline
\hline
     $k$  & $x$   & $\Delta$ $x$   &  $F_{n,k}^{(P)}(Q^{2})$  & $\Delta F_{n,k}^{(err P )}(Q^{2})$ &  Res $F_{n,k}^{(P)}(Q^{2})$  & DIS     $F_{n,k}^{(P)}(Q^{2})$  &   $F_{n,k}^{(D)}(Q^{2})$  & $\Delta F_{n,k}^{(err D )}(Q^{2})$ &  Res $F_{n,k}^{(D)}(Q^{2})$  & DIS     $F_{n,k}^{(D)}(Q^{2})$    \\
\hline
\hline
   
     0                 &        0.0588             &        0.0555             &     0.4220                 &     0.3414E-02             &        0.0000             &        1.0000             & 0.4027 & 0.3125E-02 & 0.0000 & 1.0000 \\
      1                 &        0.1176             &        0.0537             &     0.3819                 &     0.3224E-02             &        0.0000             &        1.0000             & 0.3505 & 0.2980E-02 & 0.0000 & 1.0000 \\
      2                &        0.1765             &        0.0519             &     0.3515                 &     0.2945E-02             &        0.0000             &        1.0000             & 0.3117 & 0.2815E-02 & 0.0000 & 1.0000 \\
      3                 &        0.2353             &        0.0500             &     0.3164                 &     0.2690E-02             &        0.0000             &        1.0000             & 0.2717 & 0.2632E-02 & 0.0000 & 1.0000 \\
      4                 &        0.2941             &        0.0480             &     0.2765                 &     0.2526E-02             &        0.0000             &        0.9999             & 0.2306 & 0.2459E-02 & 0.0000 & 0.9999 \\
      5                 &        0.3529             &        0.0460             &     0.2338                 &     0.2368E-02             &        0.0002             &        0.9996             & 0.1900 & 0.2248E-02 & 0.0002 & 0.9996 \\
      6                 &        0.4118             &        0.0438             &     0.1907                 &     0.2150E-02             &        0.0013             &        0.9980             & 0.1515 & 0.1988E-02 & 0.0012 & 0.9981 \\
      7                 &        0.4706             &        0.0416             &     0.1500                 &     0.1893E-02             &        0.0057             &        0.9916             & 0.1168 & 0.1722E-02 & 0.0053 & 0.9921 \\
      8                 &        0.5294             &        0.0392             &     0.1141                 &     0.1644E-02             &        0.0204             &        0.9713             & 0.8730E-01 & 0.1491E-02 & 0.0197 & 0.9724 \\
      9                 &        0.5882             &        0.0367             &     0.8445E-01             &     0.1418E-02             &        0.0607             &        0.9192             & 0.6377E-01 & 0.1292E-02 & 0.0604 & 0.9202 \\
      10                 &        0.6471             &        0.0340             &     0.6152E-01             &     0.1203E-02             &        0.1488             &        0.8131             & 0.4611E-01 & 0.1090E-02 & 0.1515 & 0.8114 \\
      11                &        0.7059             &        0.0310             &     0.4422E-01             &     0.1029E-02             &        0.2990             &        0.6456             & 0.3322E-01 & 0.8860E-03 & 0.3090 & 0.6373 \\
      12                 &        0.7647             &        0.0277             &     0.3080E-01             &     0.9280E-03             &        0.4966             &        0.4424             & 0.2349E-01 & 0.7160E-03 & 0.5143 & 0.4275 \\
      13                &        0.8235             &        0.0240             &     0.1978E-01             &     0.7860E-03             &        0.6968             &        0.2528             & 0.1552E-01 & 0.5410E-03 & 0.7168 & 0.2364 \\
      14                &        0.8824             &        0.0196             &     0.1060E-01             &     0.5460E-03             &        0.8531             &        0.1160             & 0.8891E-02 & 0.3180E-03 & 0.8719 & 0.1012 \\
      15                 &        0.9412             &        0.0139             &     0.3682E-02             &     0.3140E-03             &        0.9457             &        0.0407             & 0.4347E-02 & 0.1410E-03 & 0.9664 & 0.0253 \\

 \hline
\hline


\end{tabular}
\label{table:7P} 
\end{center}
 \end{table}
\end{widetext}

\clearpage
\bibliography{DIS}

\end{document}